\documentclass[10pt,final,letterpaper,oneside,twocolumn]{IEEEtran}

\usepackage{csquotes}

\usepackage[cmex10]{amsmath}
\usepackage{amsfonts,amssymb,amsthm}

\usepackage{mathrsfs} %
\usepackage{stmaryrd}

\usepackage{float}

\ifCLASSOPTIONcompsoc 
\usepackage[tight,normalsize,sf,SF]{subfigure} 
\else 
\usepackage[tight,footnotesize]{subfigure} 
\fi 

\usepackage{xcolor}
\usepackage{xspace}
\usepackage{url}
\usepackage{hyperref}
\usepackage[vlined,linesnumbered,noresetcount]{algorithm2e} %
\usepackage{listings}
\usepackage{graphicx}
\usepackage[draft]{fixme} %

\usepackage{code}
\usepackage{newcommands_nico}

\DontPrintSemicolon

\DeclareSymbolFont{manuscrit}{U}{manuscrit}{m}{n}
\DeclareSymbolFontAlphabet{\mathm}{manuscrit}

\newcommand{\hlinetop}{%
\kern2pt\hrule\relax \smallskip%
}

\newcommand{\hlinebot}{%
\smallskip \kern2pt\hrule\relax%
}

\newcommand{\sta}[1]{\ensuremath{\mathit{st}(#1)}\xspace}
\newcommand{\clo}[1]{\ensuremath{\mathit{cl}(#1)}\xspace}

\newcommand{\Iu}{\ensuremath{\mathfrak{I}(u)}\xspace}

\newcommand{\U}{\ensuremath{\textsc{u}}\xspace}
\newcommand{\USC}{USC\xspace}

\newcommand{\T}{\ensuremath{\mathcal{T}}\xspace}

\newcommand{\Heals}{\mathbb{H}\xspace}
\newcommand{\Keals}{\mathbb{K}\xspace}

\newcommand{\SHAPESINF}{\mathcal{S}_{\triangleleft}\xspace}
\newcommand{\SHAPESSUP}{\mathcal{S}_{\triangleright}\xspace}
\newcommand{\KHAL}[1]{\mathbb{H}^{#1}\xspace}
\newcommand{\canonicalizetree}{\textsc{canonicalize\_tree}}
\newcommand{\ub}{u^{\flat}\xspace}

\newcommand{\dom}{\mathcal{D}\xspace}
\newcommand{\JSET}{\mathcal{J}\xspace}
\newcommand{\ostar}{o^*\xspace}

\newcommand{\connectedcomponent}{\mathcal{CC}\xspace}
\newcommand{\intbd}{\partial_{\mathrm{int}}\xspace}
\newcommand{\CLOSEDDOM}{\bar{\dom}\xspace}

\newcommand{\shape}{\mathcal{S}\xspace}
\newcommand{\level}{\mathscr{L}\xspace}

\newcommand{\front}{P\xspace}
\newcommand{\cav}{\T\xspace}
\newcommand{\cavi}[1]{\cav_{#1}\xspace}
\newcommand{\cavij}[2]{\cav_{#1}^{#2}\xspace}

\newcommand{\levelij}[2]{\level_{#1}^{#2}\xspace}
\newcommand{\leveli}[1]{\level_{#1}\xspace}

\newcommand{\HierarchyCavities}{\mathfrak{PR}\text{-}\mathfrak{Cav}\xspace}
\newcommand{\parentCAV}{\textbf{parent}_{\HierarchyCavities}\xspace}
\newcommand{\nextCAV}{\textbf{next}_{\HierarchyCavities}\xspace}
\newcommand{\properCAV}[1]{\left(#1\right)_{*,\HierarchyCavities}\xspace}
\newcommand{\levelCAV}{\level_{*,\HierarchyCavities}\xspace}
\newcommand{\childrenCAVITY}{\textbf{children}_{\HierarchyCavities}\xspace}

\newcommand{\ToS}{\mathfrak{S}\xspace}
\newcommand{\nextSHAPE}{\textbf{next}_{\ToS}\xspace}
\newcommand{\childrenSHAPE}{\textbf{children}_{\ToS}\xspace}
\newcommand{\properSHAPE}[1]{\left(#1\right)_{*,\ToS}\xspace}
\newcommand{\parentSHAPE}{\textbf{parent}_{\ToS}\xspace}
\newcommand{\levelSHAPE}{\level_{*,\ToS}\xspace}

\newcommand{\bd}{\partial \xspace}

\newcommand{\Interior}{\mathrm{Int}\xspace}

\newcommand{\Cl}[1]{\ensuremath{\mathit{cl}_{\CLOSEDDOM}({#1})}\xspace}
\newcommand{\St}[1]{\ensuremath{\mathit{st}_{\CLOSEDDOM}({#1})}\xspace}
\newcommand{\lpar}{\mathscr{L}_{par}\xspace}
\newcommand{\PROPAG}{P\xspace}

\newcommand{\border}{\triangle\xspace}

\newcommand{\jstar}{j^*\xspace}
\newcommand{\PARENT}{\cavij{\istar}{\jstar}\xspace}
\newcommand{\istarstar}{i^{**}\xspace}
\newcommand{\jstarstar}{j^{**}\xspace}
\newcommand{\PARENTDEUX}{\cavij{\istarstar}{\jstarstar}\xspace}

\newcommand{\EspaceInterieur}{\mathfrak{I}\xspace}
\newcommand{\EspaceExterieur}{\mathfrak{E}\xspace}
\newcommand{\BOUNDARY}{bd\xspace}
\newcommand{\istar}{i^*\xspace}
\newcommand{\BOTH}{\mathscr{F}\xspace}
\newcommand{\PARENTSTRUCTURE}{\mathit{parent}_{UF}\xspace}

\newcommand{\SBIPMAP}{$\mathcal{SI}$-map\xspace}
\newcommand{\SBIPMAPS}{$\mathcal{SI}$-maps\xspace}
\newcommand{\SPAN}{\mathrm{span}\xspace}

\newcommand{\WC}{well-com\-po\-sed\xspace}

\newtheorem{Notation}{Notation}
\newtheorem{Property}{Property}
\newtheorem{Remark}{Remark}
\newtheorem{Definition}{Definition}

\newtheorem{Proposition}{Proposition}
\newtheorem{Coro}{Corollary}
\newtheorem{Theorem}{Theorem}

\makeatletter
\hypersetup{pdftitle=\@title,
  pdfauthor={Thierry G\'eraud, Nicolas Boutry, S\'ebastien Crozet, Edwin Carlinet, Laurent Najman},
  pdfkeywords={Mathematical Morphology}}
\makeatother

\begin{document}

\title{A Proof of the Tree of Shapes in $n$-D}
\author{Thierry G\'eraud, %
  Nicolas Boutry$*$, %
  S\'ebastien Crozet, %
  Edwin Carlinet, %
  Laurent Najman
\thanks{$*$ E-mail: \tt nicolas.boutry@lrde.epita.fr}
 \thanks{M.\ G\'eraud and M.\ Boutry are with
    the EPITA Research and Development Laboratory (LRDE),
    14-16, rue Voltaire, FR-94276 Le Kremlin-Bic\^etre Cedex, France.}%
 \thanks{M.\ Najman is with
    Universit\'e Gustave-Eiffel, Laboratoire d'Informatique Gaspard-Monge,
    \'Equipe A3SI, ESIEE Paris, Cit\'e Descartes, BP 99, FR-93162
    Noisy-le-Grand Cedex, France}%
 \thanks{E-mails: firstname.lastname@lrde.epita.fr, laurent.najman@esiee.fr}
}

\makeatletter
\markboth{}%
{G\'eraud \MakeLowercase{\textit{et al.}}: \@title}
\makeatother

\maketitle

\begin{abstract}
In this paper, we prove that the self-dual morphological hierarchical structure computed on a $n$-D gray-level well-composed image $u$ by the algorithm of Géraud \textit{et al.}~\cite{geraud.2013.ismm} is exactly the mathematical structure defined to be the \emph{tree of shape} of $u$ in Najman \textit{et al}~\cite{najman.2013.ismm}. We recall that this algorithm is in quasi-linear time and thus considered to be \emph{optimal}. The tree of shapes leads to many applications in mathematical morphology and in image processing like grain filtering, shapings, image segmentation, and so on.
\end{abstract}


\section{Introduction}
\label{sec:introduction}

The material presented here is a formal proof that the hierarchical structure provided by the main algorithm in~\cite{geraud.2013.ismm} is the tree of shapes presented in~\cite{najman.2013.ismm}. This proof is $n$-dimensional.

\section{Mathematical background}
\label{sec:background}

In the following, we consider a \nD digital image $u$ as a function
defined on a regular cubical \nD grid and having integral values (even if we can generalize our case to any totally ordered set); in brief, we have then $u : \Zeals^n \rightarrow \Zeals$.

Practically a digital image is defined on a finite domain, usually an hyper-rectangle $\set{(z_1,
  \ldots, z_n)}{\forall i,\, 0 \leq z_i < N_i}$, so the number of
points is $N = \prod_{i=1}^{n} N_i$, and the space of values is
restricted to $\llbracket\, 0, \, 2^Q - 1 \,\rrbracket$, where
$Q$ is the quantization.  We are interested in computing the tree of
shapes of such images.

For generality purpose, we give below some definitions related to a
function $f$ defined upon a discrete set $X$ and taking values in a
finite set $Y$; we have $f: X \rightarrow Y$.  With $X$ having some
discrete topology, for any subset $E \subset X$, we denote by $\CC(E)$
the set of connected components of $E$.  Given $x \in X$, we denote by
$\CC(E,x)$ the connected component of $E$ containing $x$ if $x \in E$; otherwise, we set $\CC(E,x) := \emptyset$.

\newcommand{\ZN}{\Zeals^n\xspace}
\newcommand{\Neals}{\mathbb{N}\xspace}

\newcommand{\ZNSURS}{\frac{1}{s} \ZN\xspace}

\subsection{The tree of shapes in $\Zeals^n$}

Let us now consider $X$ as being defined on the regular cubical \nD grid $\ZN$ or on one of its subdivisions like $\ZNSURS$ with $ s \in \Neals^*$. Furthermore, $X$ is \emph{unicoherent} in the sense that it is connected and for any two subsets $A,B$ ox $X$, the set $A \cap B$ is connected.

\medskip

As usual in discrete topology, to properly deal with subsets of $\ZN$ and with their complementary in $\ZN$, we consider the dual connectivities $\cmin$ and $\cmax$. 

\medskip

The extension from set to gray-level images is done using what are called threshold sets. For any $\lambda \in \Reals$, the subsets of $X$:
\begin{align*}
  \mincut{u} &\,:=\, \{\, x \in X \;|\; u(x) < \lambda \,\}, \\
  \text{and~} \maxcut{u} &\,:=\, \{\, x \in X \;|\; u(x) \geq \lambda \,\}
\end{align*}
are respectively called the \emph{strict lower threshold set} and \emph{large upper threshold set} of the function $f$ relatively to (the threshold) $\lambda$. The sets $\mincut{u}$ and \maxcut{u} can be understood as binary images.

From these two families of threshold sets, we can deduce two sets, $\Tmin(u)$ and $\Tmax(u)$, composed of the connected components of respectively lower and upper cuts of $u$:
\begin{align*}
\Tmin(u) &\,=\, \{\, \Gamma \in \CCmin(\mincut{u}) \,\}_\lambda, \\
\Tmax(u) &\,=\, \{\, \Gamma \in \CCmax(\maxcut{u}) \,\}_\lambda.
\end{align*}
The elements of $\Tmin(u)$ and $\Tmax(u)$ respectively give rise to two dual trees\footnote{We say that the min-tree and the max-tree are \emph{dual} in the sense that for any image $u$, the structure of the min-tree of $u$ is equal to the structure of the max-tree of $-u$.}: the min-tree and the max-tree of $u$.

Let us recall what we call a \emph{shape} but before let us reintroduce some mathematical basics for mathematical morphology. The \emph{saturation operator} fills in the cavities of subsets of a topological space $\Omega$ this way for any $\Gamma \subseteq \Omega$:
$$\sat(\Gamma) = \Omega \setminus \CC(\Omega \setminus \Gamma,\pinfty).$$
In discrete topology, we obtain then for any $\Gamma \subseteq \ZN$:
\begin{align*}
\sat_\cmax(\Gamma) &\,:=\, \ZN \setminus \CCmax(\ZN \setminus \Gamma),\\
\sat_\cmin(\Gamma) &\,:=\, \ZN \setminus \CCmin(\ZN \setminus \Gamma).
\end{align*}

Based on these trees, we can define what we call the \emph{sets of shapes}: 
\begin{align*}
\Smin(u) &\,:=\, \{\, \sat_\cmax(\Gamma); \; \Gamma \in \Tmin(u) \,\} \\
\Smax(u) &\,:=\, \{\, \sat_\cmin(\Gamma); \; \Gamma \in \Tmax(u) \,\}
\end{align*}

$\Smin(u)$ is the set of lower shapes of $u$, $\Smax(u)$ is the set of upper shapes of $u$. They correspond to the sets of elements of the min- and max-trees when we have filled in their cavities using the saturation operator.

For a given subset $\Gamma \subseteq \Omega$, and for a arbitrarily chosen element $\pinfty \in \Omega$, then we call \emph{cavity} each connected component of $\Omega \setminus \Gamma$ and which does not contain $\pinfty$. The connected component of $\Omega \setminus \Gamma$ containing $\pinfty$ is called the \emph{exterior} of $\Gamma$ and can be empty.

\begin{figure}[!t]
\centering
\subfigure[Image $u$]{\includegraphics[width=.3\linewidth]{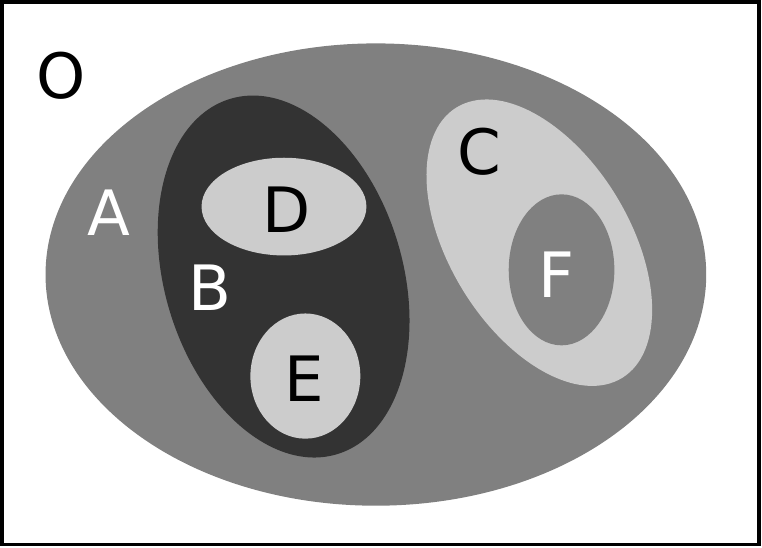}
\label{fig.dummy.image}}
\hfil
\subfigure[Gray scale]{\includegraphics[width=.3\linewidth]{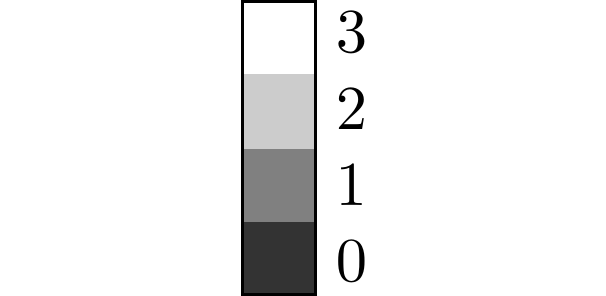}
\label{fig.dummy.scale}}
\hfil
\subfigure[Level lines of $u$]{\includegraphics[width=.3\linewidth]{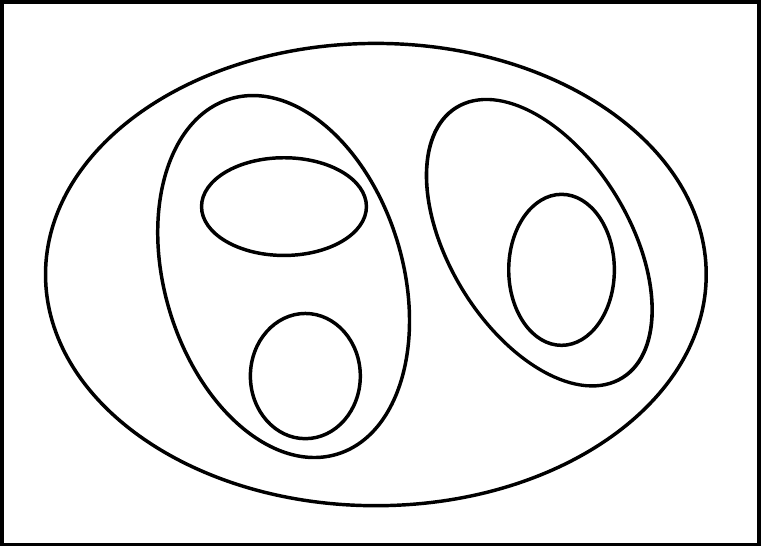}
\label{fig.dummy.scale}}

\subfigure[${[\,u < 2\,]}$ \label{fig:dummy.lower2}]{\includegraphics[width=.3\linewidth]{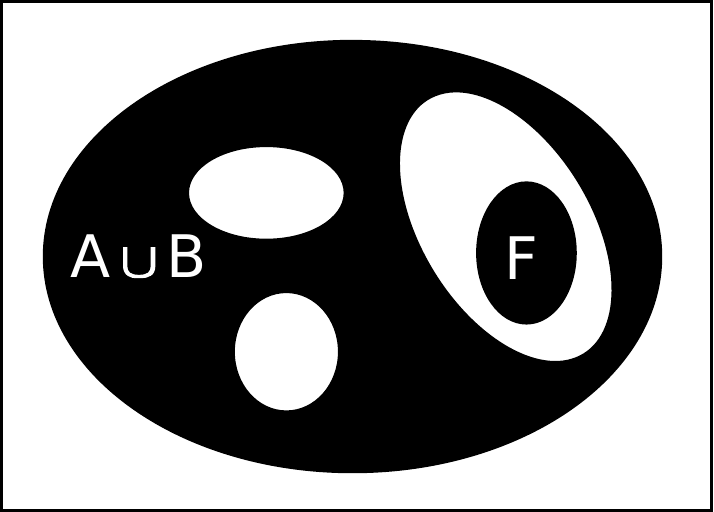}}
\hfil
\subfigure[${[\,u \geq 1\,]}$ \label{fig:dummy.upper1}]{\includegraphics[width=.3\linewidth]{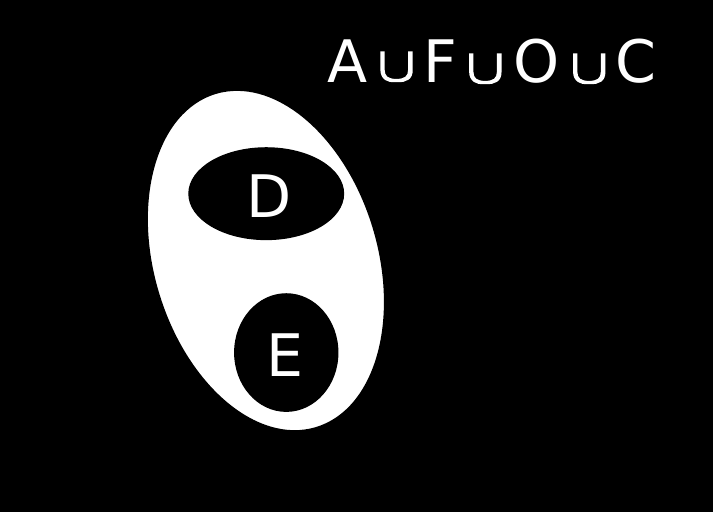}}
\hfil
\subfigure[A shape from (d) \label{fig:dummy.shape}]{\includegraphics[width=.3\linewidth]{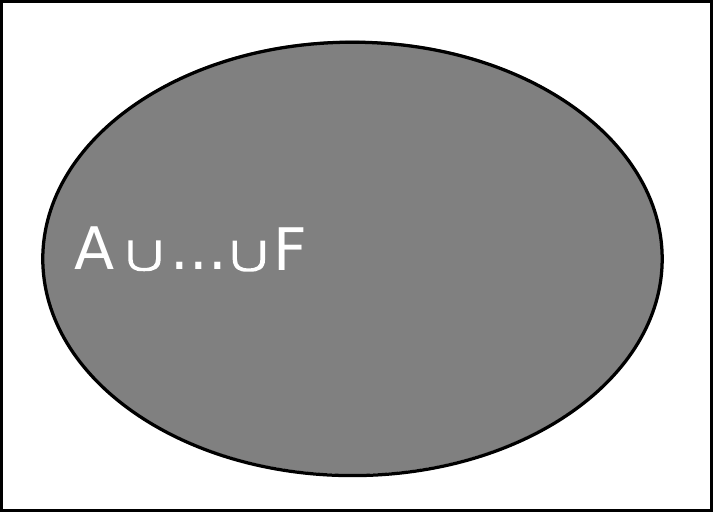}}

\vskip-0.5em

\subfigure[${[\,u < 1\,]}$ \label{fig:dummy.lower1}]{\includegraphics[width=.3\linewidth]{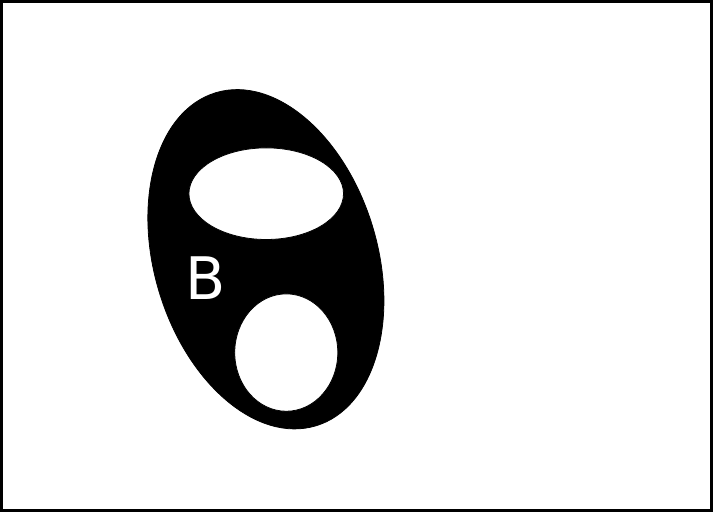}}
\hfil
\subfigure[${[\,u \geq 2\,]}$ \label{fig:dummy.upper2}]{\includegraphics[width=.3\linewidth]{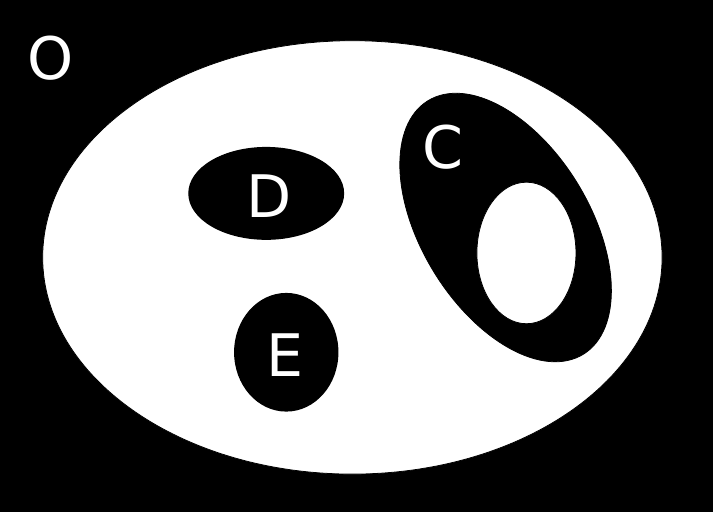}}
\hfil
\subfigure[Two shapes resp. $~~~~~~$from (g) and (h)\label{fig:dummy.shapes}]{\includegraphics[width=.3\linewidth]{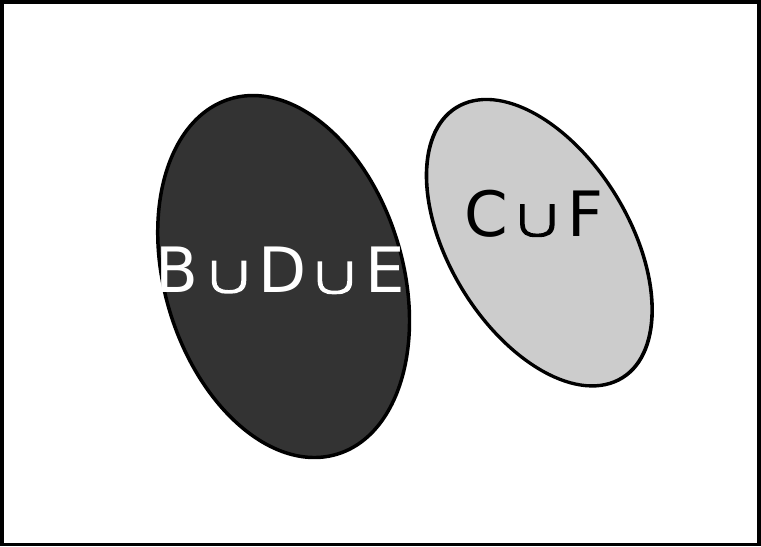}}

\subfigure[Min-tree $\Tmin(u)$]{\includegraphics[scale=.35]{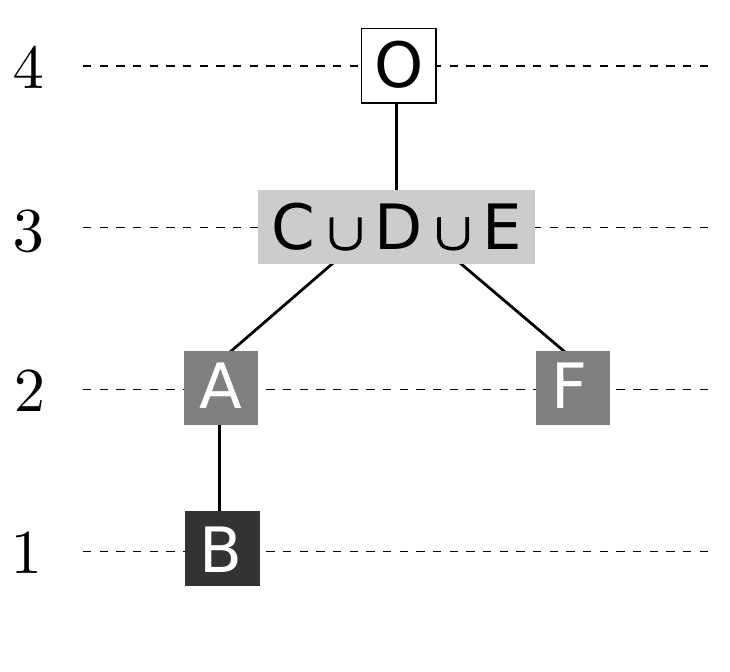}
\label{fig.dummy.mintree}}
\hfil
\subfigure[Max-tree $\Tmax(u)$]{\includegraphics[scale=.35]{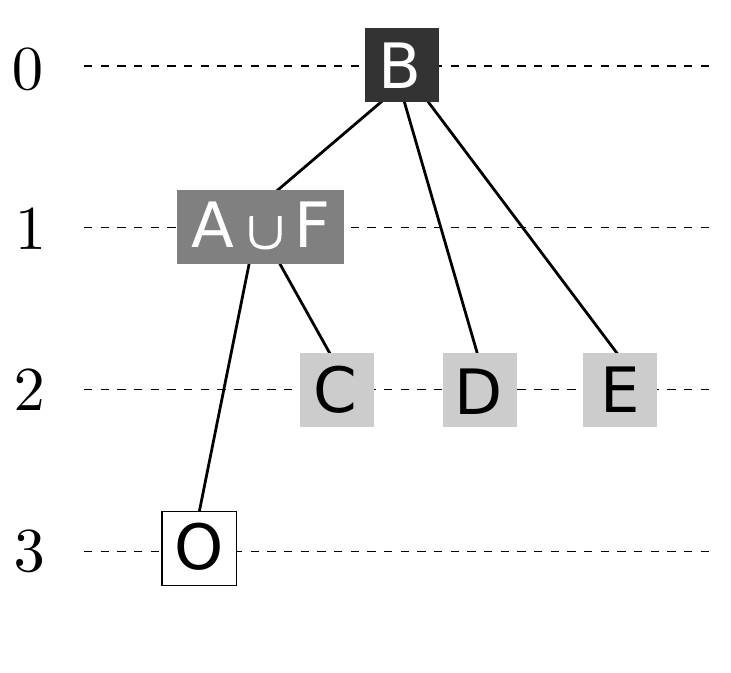}
\label{fig.dummy.maxtree}}
\hfil
\subfigure[Tree of shapes $\Tos(u)$]{\includegraphics[scale=.35]{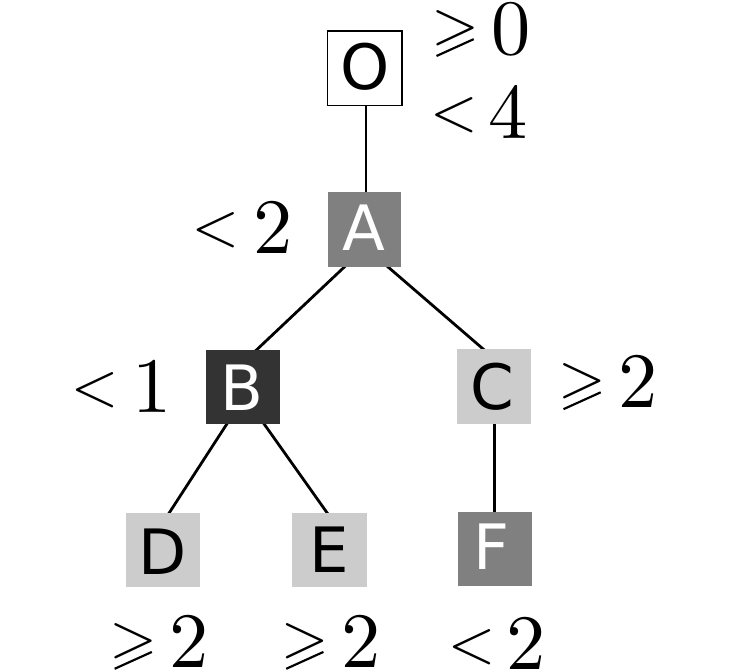}
\label{fig.dummy.tos}}

\caption{Morphological trees.}
\label{fig.dummy}
\end{figure}

The set of all shapes:
\begin{align}
\Tos(u) \,=\, \Smin(u) \,\cup\, \Smax(u)
\end{align}
forms a tree, the so-called \textit{tree of shapes} of $u$~\cite{monasse.2000.itip}.  Indeed, for any pair of shapes $\Gamma$ and $\Gamma'$ in $\Tos$, we have $\Gamma \subset \Gamma' \mbox{~~or~~} \Gamma' \subset \Gamma \mbox{~~or~~} \Gamma \cap \Gamma' = \emptyset$.

Actually, the shapes are the cavities of the elements of $\Tmin$ and $\Tmax$.  For instance, if we consider a lower component $\Gamma \in \mincut{f}$ and a cavity $H$ of $\Gamma$, this cavity is an upper shape, i.e., $H \in \Smax$.  Furthermore, in a discrete setting, $H$ is obtained after having filled the cavities of a component of $\maxcut{f}$.

Figure~\ref{fig.dummy} depicts on a sample image the three components trees ($\Tmin$, $\Tmax$, and $\Tos$).  Just note that the
Equations so far rely on the pair of dual connectivities, \cmin and \cmax, so discrete topological problems are avoided, and, in addition, we are forced to consider two kind of cuts: strict ones for \cmin and large ones for \cmax.

\subsection{Cellular complex and {K}halimsky grid}
\label{ssec:cellular.complex}

\newcommand{\starHN}{st_{\KHAL{n}}\xspace}
\newcommand{\cloHN}{cl_{\KHAL{n}}\xspace}

Let us recall the definitions relative to \nD Khalimsky grids~\cite{khalimsky1990computer,mazo.2012.jmiv.a}. From the sets $\KHAL{1}_0=\, \{ \{a\} ; a \in \Zeals \}\,$ and
$\KHAL{1}_1 \,=\, \{ \{a,a+1\} ; a \in \Zeals \}$, we can define $\KHAL{1}
\,=\, \KHAL{1}_0 \cup \KHAL{1}_1\,$ and the set $\KHAL{n}$ as the $n$-ary Cartesian
power of $\KHAL{1}$.  If an element $h \subset \Zeals^n$ is the Cartesian
product of $d$ elements of $\KHAL{1}_1$ and $(n - d)$ elements of $\KHAL{1}_0$, we
say that $h$ is a $d$-face of $\KHAL{n}$ and that $d$ is the dimension of
$h$.  The set of all faces, $\KHAL{n}$, is called the \emph{\nD space of cubical
complexes}.  Figure~\ref{fig:E} depicts a set of faces $\{f,g,h\}
\subset \KHAL{2}$ where $f=\tset{0}\ttimes\tset{1}$,
$g=\tset{0,1}\ttimes\tset{0,1}$, and $h=\tset{1}\ttimes\tset{0,1}$;
the dimension of those faces are respectively 0, 2, and 1.  Let us
write $\starHN(h) = \{ h' \in \KHAL{n} \,|\, h \subseteq h' \}$ and
$\cloHN(h) = \{ h' \in H^n \,|\, h' \subseteq h \}$.  The pair
$(H^n,\subseteq)$ forms a poset and the set $\, \mathcal{U} = \{ U
\subseteq H^n \,|\, \forall h \in U, \, \starHN(h) \subseteq U \}$ is
a {T0-Alexandroff} topology on $\KHAL{n}$.  With $E \subseteq \KHAL{n}$, we have
a star operator $\sta{E} = \cup_{h \in E}\, \starHN(h)$ and a closure
operator $\clo{E} = \cup_{h \in E}\, \cloHN(h)$, that respectively
gives the smallest open set and the smallest closed set of
$\power{\KHAL{n}}$ containing $E$.

\begin{figure}[t]
  \centering
  \subfigure[In $\power{\Zeals^2}$. \label{fig:faces.Z2}]{
    \includegraphics[width=.22\linewidth]{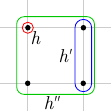}}
  \hfil
  \subfigure[In $\Keals^2$. \label{fig:faces.K2}]{
    \includegraphics[width=.22\linewidth]{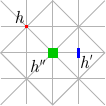}}
  \hfil
  \subfigure[In $\power{\Reals^2}$. \label{fig:faces.R2}]{
    \includegraphics[width=.22\linewidth]{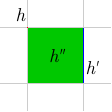}}
  \hfil
  \subfigure[In $\Heals^2$. \label{fig:faces.H2}]{
    \includegraphics[width=.22\linewidth]{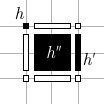}}
  \caption{Three faces, $h, h'$ and $h''$, depicted as subsets of
    $\Zeals^2$ (a), as vertices of the Khalimsky grid (b), as
    geometrical objects, parts of the plane (c), and as elements of
    the cellular complex $\Heals^2$ (d). \label{fig:faces}
  }
\end{figure}

The set of faces of $\KHAL{n}$ is arranged onto a grid, the so-called
\emph{Khalimsky's grid}, depicted in gray in
Figure~\ref{fig:faces.K2}.  The set of $n$-faces is denoted by $\KHAL{n}_n$; it is the
$n$-Cartesian product of $\KHAL{1}_1$.

\begin{figure}[t]
  \centering
  \subfigure[$E \subset \Heals^2$ \label{fig:E.E}]{
    \includegraphics[width=.20\linewidth]{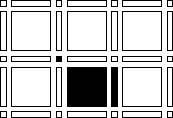}}
  \hfil
  \subfigure[$\sta{E}$ \label{fig:E.sta}]{
    \includegraphics[width=.20\linewidth]{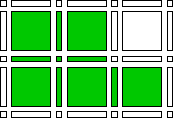}}
  \hfil
  \subfigure[$\clo{E}$ \label{fig:E.clo}]{
    \includegraphics[width=.20\linewidth]{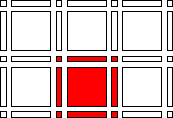}}
  
  \subfigure[$\mathit{int}(E)$ \label{fig:E.int}]{
    \includegraphics[width=.20\linewidth]{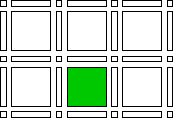}}
  \hfil
  \subfigure[$\partial E$ \label{fig:E.boundary}]{
    \includegraphics[width=.20\linewidth]{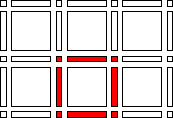}}
  \hfil
  \subfigure[$\mathit{ext}(E)$ \label{fig:E.ext}]{
    \includegraphics[width=.20\linewidth]{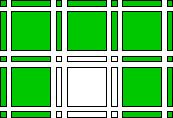}}
  \caption{Some topological operators on subsets of
    $\KHAL{n}$. \label{fig:E} }
\end{figure}

\subsection{Set-Valued Maps}
\label{ssec:set.valued.maps}

Now let us recall the mathematical background relative to set-valued maps~\cite{aubin.2008.book}. A set-valued map $~\U \,:\, X \, \leadsto \,Y\;$ is characterized by
its graph, $\,\Gra(\U) \isset{ (x, y) \in X \times Y }{ y \in \U(x)
}.$ There are two different ways to define the ``inverse'' of a subset
by a set-valued map: $\, \U^\oplus(M) \;=\; \set{ x \in X }{ \U(x)
  \cap M \neq \emptyset } \,$ is the \textit{inverse image} of $M$ by
$\U$, whereas $\, \U^\ominus(M) \;=\; \set{ x \in X }{ \U(x) \subset M
} \,$ is the \textit{core} of $M$ by $\U$. Two distinct continuities
are defined on set-valued maps.  The one we are interested in is the
``natural'' extension of the continuity of a single-valued function.
When $X$ and $Y$ are metric spaces and when $\,\U(x)\,$ is compact,
$\,\U\,$ is said to be \textsc{u}pper
\textsc{s}emi-\textsc{c}ontinuous (\USC) at $\,x\,$ if $\, \forall
\varepsilon > 0, \; \exists \, \eta > 0 \,$ such that $\, \forall \,
x' \in B_X(x,\eta), \; \U(x') \subset B_Y(\U(x),\varepsilon), \,$
where $B_X(x,\eta)$ denotes the ball of $X$ of radius $\eta$ centered
at $x$.  One characterization of \USC maps is the following: $\,\U\,$
is \USC if and only if the core of any open subset is open.

\begin{figure}[bt]
  \centering
  \subfigure[Image $u : \Zeals^2 \rightarrow \Zeals$ \label{fig:surf.ima}]{
    ~~~~~\includegraphics[width=.15\linewidth]{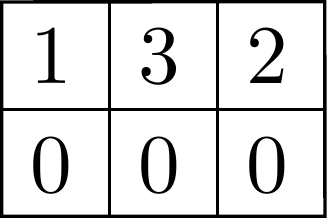}~~~~~}
  \hfil
  \subfigure[Surface of $u$ \label{fig:surf.surf}]{
    ~~~~~~\includegraphics[width=.30\linewidth]{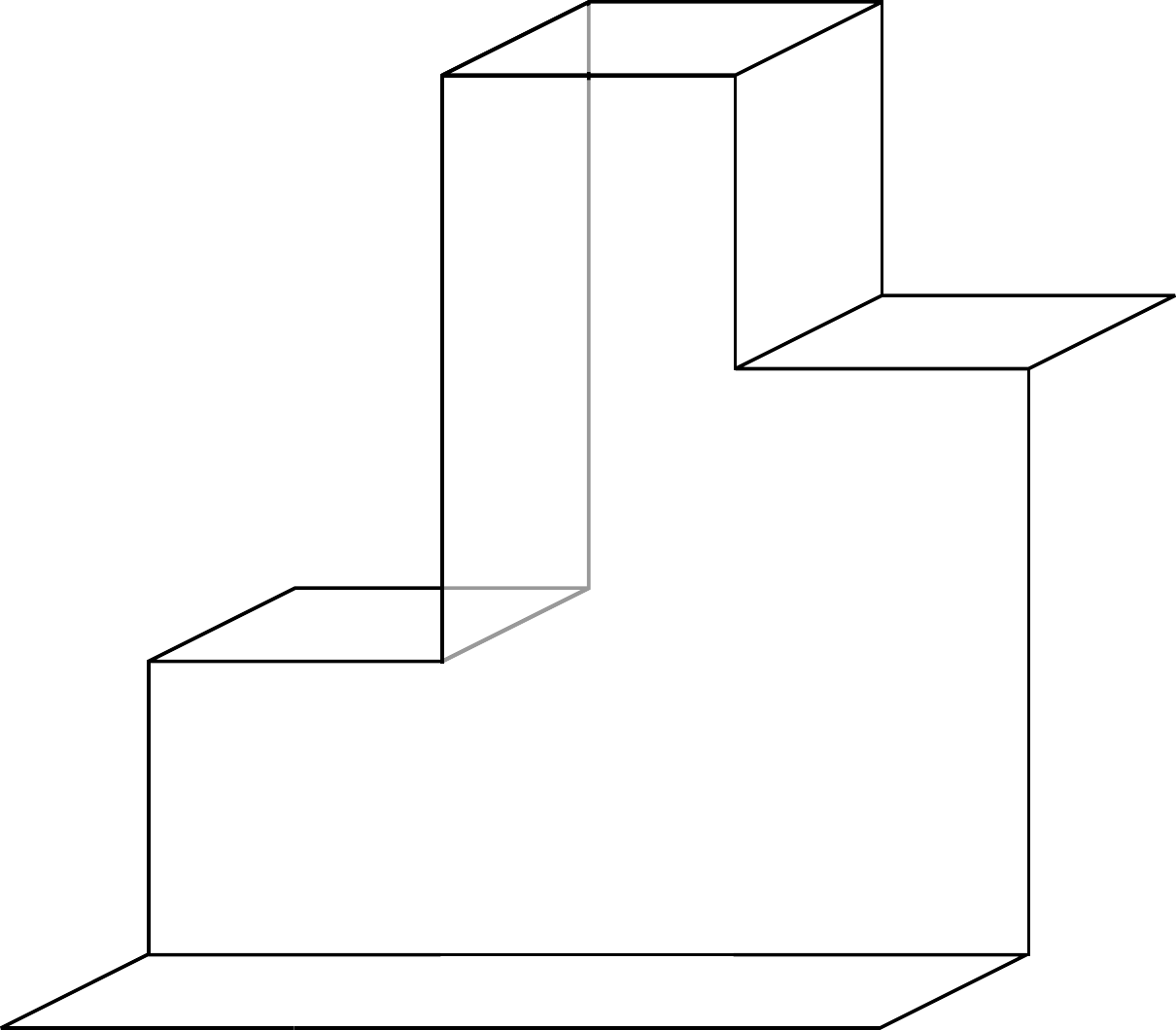}}
  
  \subfigure[Image $U : \Heals^2 \rightarrow \mathbb{I}_\Zeals$ \label{fig:surf.H}]{
    \includegraphics[width=.45\linewidth]{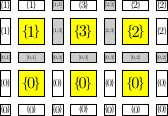}}
  \hfil
  \subfigure[$U$ seen in 3D \label{fig:surf.H3D}]{
    \includegraphics[width=.45\linewidth]{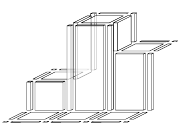}}
\caption{From the classical view to the topographical view: In (a), a 2D image $u$ made of six pixels. In (b), it topographical view (where discontinuities can be observed).  In (c), the \USC span-based immersion $U$ of $u$ where the 2-faces (elements of $\KHAL{2}_2$) are depicted in yellow; they correspond to the pixels of $u$. In (d), the topographical view of the \USC map $U$ without discontinuities (in the sense of an \USC map).}
\label{fig:surf}
\end{figure}

A topographical view of a 2D image can be observed in Figure~\ref{fig:surf} and shows that, using span-based immersion maps, we obtain \textquote{continuous} maps (in the sense of set-valued maps).

\subsection{Interpolation}
\label{ssec:interpolation}

We are going to immerse a discrete \nD function defined on a cubical grid $\,u :
\Zeals^n \rightarrow \Zeals\,$ into some larger spaces in order to get
some continuity properties.  For the \textit{domain space}, we use the
subdivision $X = \half \KHAL{n}$ of $\KHAL{n}$.  Every element $z \in \Zeals^n$
is mapped to an element $\,m(z) \in \half \KHAL{n}_n\,$ of dimension $n$ with $z =
(z_1,\ldots,z_n) ~ \longmapsto ~ m(z) \, = \, \{z_1,z_1+\half\} \times
\ldots \times \{z_n,z_n+\half\}$. The definition domain of $u$,
$\mathcal{D} \subseteq \Zeals^n$, has thus a counterpart in $X$, that
will also be denoted $\mathcal{D}$, and that is depicted in bold in
Figure~\ref{fig.intermax}.

For the \textit{value space}, we immerse
$\Zeals$ (the set of pixel values) into the larger space $Y = \half
\KHAL{1}$, where every integer becomes a closed singleton, that is, an element of $\KHAL{1}_0$.
Thanks to an \textquote{interpolation} function, we can now define from $u$ a
set-valued map $\U = \Iu$.  We have $~\U \,:\, X \, \leadsto \,Y\,$
and we set $  \forall\, h \in X, \;\; \U(h) \,:= \, $
\begin{equation*}
  \label{eq.interop}
\left\{
    \begin{array}{ll}
      \{\, u(m^{-1}(h)) \, \} & \mbox{~~if ~} h \in \dom \\{}
      \max(\, \U(h') : \, h' \in \sta{\clo{h}} \cap\dom \,) & \mbox{~~if ~} h \in \half \KHAL{n}_1 \backslash \dom  \\{}
      \SPAN(\, \U(h') : \, h' \in \sta{h} \cap \dom \,) & \mbox{~~if ~} h \in X \backslash \half \KHAL{n}_1.
    \end{array}
    \right.
\end{equation*}

\begin{figure}[htbp]
  \centering
\subfigure[Primary image $u$.]{~~~~~~\includegraphics[scale=.4]{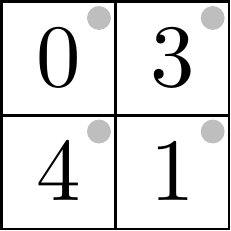}~~~~~~
\label{fig.example.u}}
\hfil
\subfigure[Subdivided image $u_s$.]{~~~~~~\includegraphics[scale=.4]{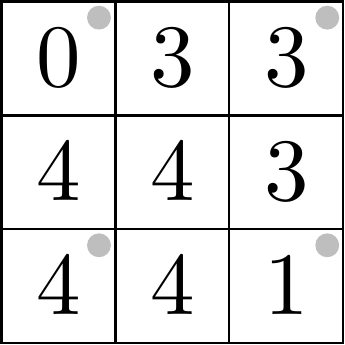}~~~~~~
\label{fig.example.us}}
\hfil
\subfigure[Immersed image $\U$.]{~~\includegraphics[scale=.4]{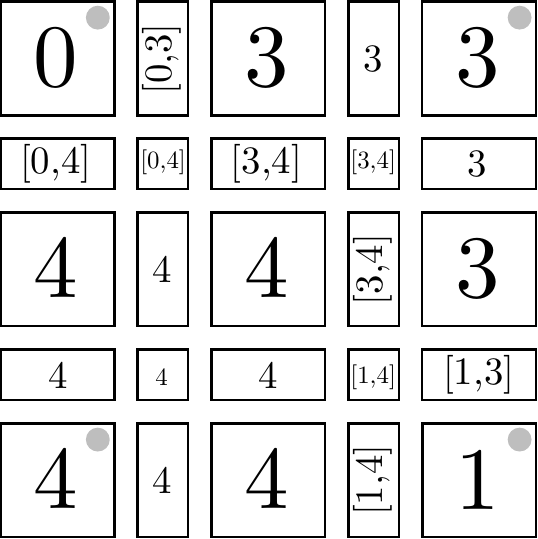}~~
\label{fig.example.uk}}
\caption{A grayscale 2D image $u$ which is not well-composed since any value in $[0,1]$ is lower than any value in $[3,4]$,  a representation $u_s$ of $u$ which is well-composed thanks to the max-interpolation, and the corresponding well-composed immersion $U$ of $u_s$ into $\KHAL{2}$.} 
\label{fig.exampleofinterpolation}
\end{figure}

\newcommand{\sclexu}{.3}

\begin{figure}[htbp]
  \centering
\begin{tabular}{cccc}
\includegraphics[scale=\sclexu]{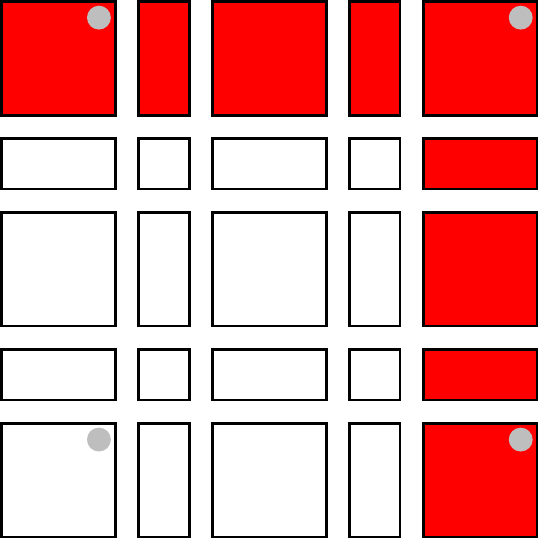} &
\includegraphics[scale=\sclexu]{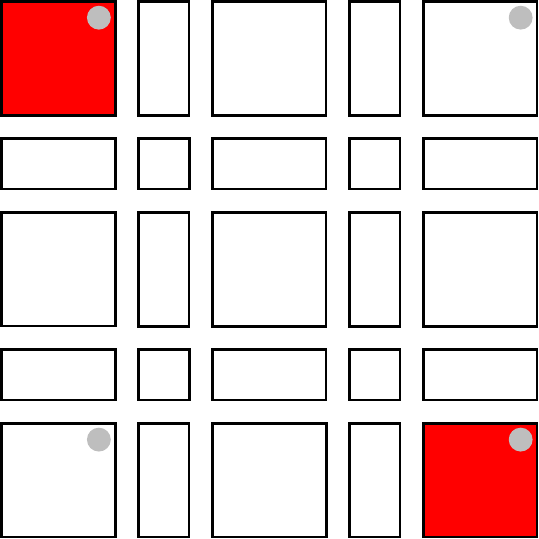} &
\includegraphics[scale=\sclexu]{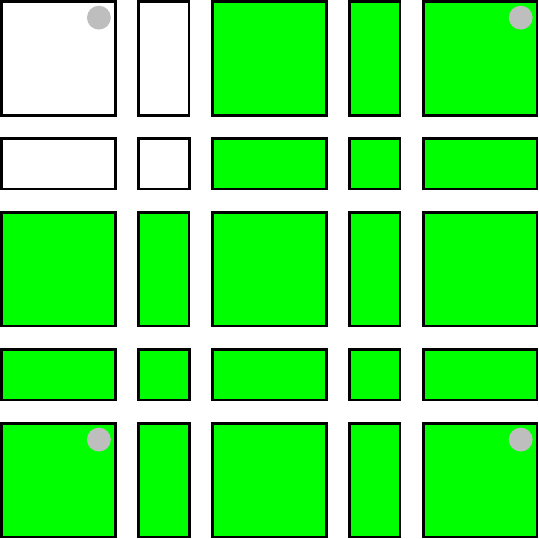} &
\includegraphics[scale=\sclexu]{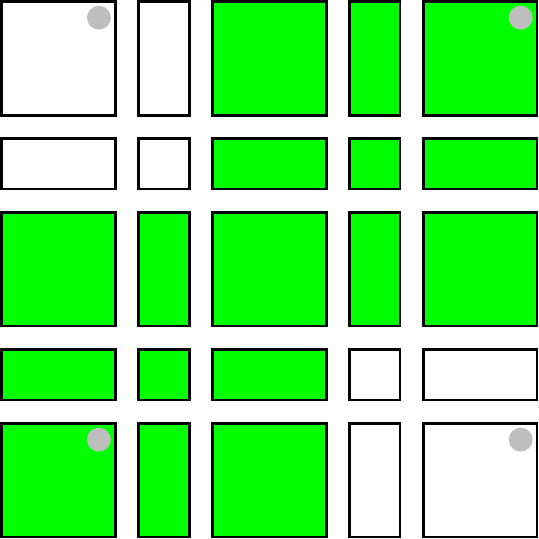} \\
\scriptsize $\lcut{\U}{4 \!-\! \epsilon}$ &
\scriptsize $\lcut{\U}{2 \!-\! \epsilon}$ &
\scriptsize $\gcut{\U}{1 \!-\! \epsilon}$ &
\scriptsize $\gcut{\U}{3 \!-\! \epsilon}$ \\
&&&\\

\includegraphics[scale=\sclexu]{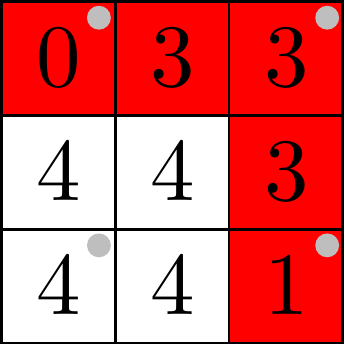} &
\includegraphics[scale=\sclexu]{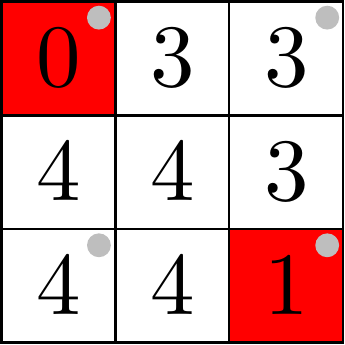} &
\includegraphics[scale=\sclexu]{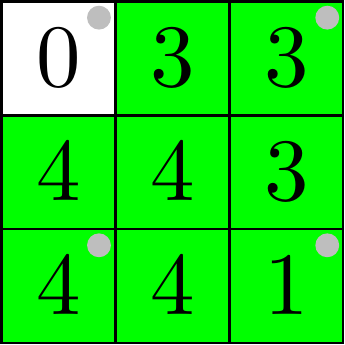} &
\includegraphics[scale=\sclexu]{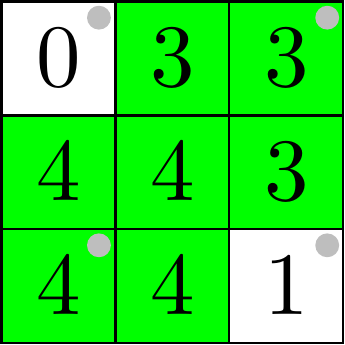} \\
\scriptsize $[u_s < 4]$ &
\scriptsize $[u_s < 2]$ &
\scriptsize $[u_s \geq 1]$ &
\scriptsize $[u_s \geq 3]$ \\
&&&\\

\includegraphics[scale=\sclexu]{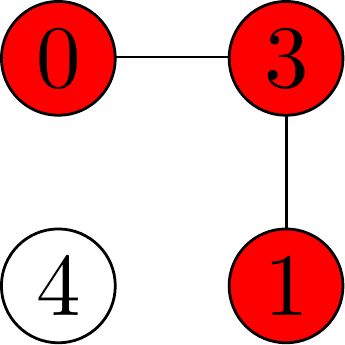} &
\includegraphics[scale=\sclexu]{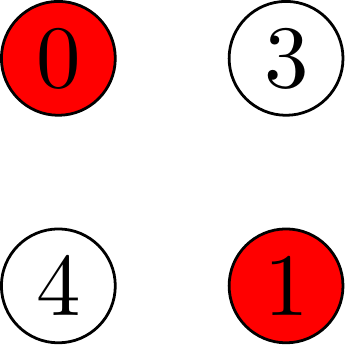} &
\includegraphics[scale=\sclexu]{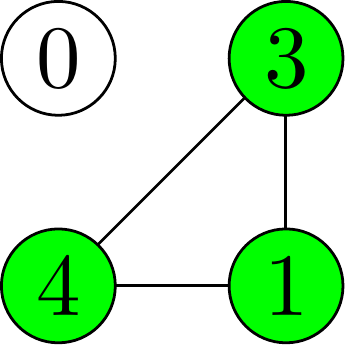} &
\includegraphics[scale=\sclexu]{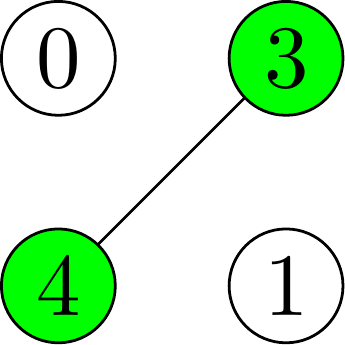} \\
\scriptsize $[u < 4]$ &
\scriptsize $[u < 2]$ &
\scriptsize $[u \geq 1]$ &
\scriptsize $[u \geq 3]$ \\
&&&\\

\end{tabular}

\caption{Since we have used the max-interpolation to make $u$ \WC, the lower threshold sets of the immersion $U$ are connected, or equivalently the lower threshold sets of $u_s$ are connected, iff the lower threshold sets of $u$ are $2n$-connected. Conversely, the upper threshold sets of $U$ are connected, or equivalently the upper threshold sets of $u_s$ are connected, iff the threshold sets of $u$ are $(3^n-1)$-connected. We have chosen here $\epsilon = \half$ because we work with integral values. Recall that in \WC images, $2n$- and $(3^n-1)$-connectivities are equivalent~\cite{boutry.15.ismm}.}
\label{fig.intermax}
\end{figure}

An example of interpolation is given in Figure~\ref{fig.exampleofinterpolation}.
Actually, whatever $u$, such a discrete interpolation
$\Iu$ can also be interpreted as a non-discrete set-valued
map $\mathfrak{I}_\Reals(u): \Reals^n \leadsto \Reals$ (schematically
$\mathfrak{I}_\Reals(u)(x) = \Iu(h)$ with $h$ such as $x
\in \Reals^n$ falls in $h \in \half \KHAL{n}$), and we can show that
$\mathfrak{I}_\Reals(u)$ is an \emph{upper semi-continuous}~\cite{aubin.2008.book} (shortly \USC) map.  

\medskip

Such span-based immersions will be called \emph{\SBIPMAPS} in the sequel.

\subsection{Discrete surfaces and well-composedness}

\newcommand{\INTERP}{\mathfrak{I}\xspace}

\newcommand{\clY}{cl_Y\xspace}
\newcommand{\stY}{st_Y\xspace}

Let us recall the definition of \emph{discrete surfaces}~\cite{najman2013discrete}. Let $Y$ be some partially ordered set (shortly poset) of rank $n \geq 0$. Let $X$ be some subset of $Y$. We say that $X$ is a discrete $(-1)$-surface when $X = \emptyset$. We say $X$ is a $0$-surface if it can be written $\{h,h'\}$ with $h \not \in \stY(h') \cup \clY(h')$. We call $X$ a $k$-surface with $k \geq 1$ when it is connected and when for any $h \in X$, the poset $\stY(h) \cup \clY(h) \setminus \{h\}$ is a discrete $(k-1)$-surface.

Furthermore, we call \emph{(combinatorial) boundary} of a subset $Z$ of $Y$ the set:
$$\partial Z := \clY(Z) \cap \clY(Y \setminus Z).$$

We say that $Z$ is \emph{(Alexandrov) well-composed} (shortly WC) when its boundary is a (possibly empty) separated union of discrete $(n-1)$-surfaces.

\begin{Remark}
We recall that the shapes of a WC \SBIPMAP are WC too.
\label{remark.shapes.WC}
\end{Remark}

\subsection{The tree of shapes in $\KHAL{n}$}

Assuming that the domain $X$ of the interpolation $\INTERP(u)$ is \emph{unicoherent}\footnote{A topological space is \emph{unicoherent}~\cite{caselles2009geometric} if it is connected, and for any two closed connected subsets $A,B$ of this domain whose union covers this same domain, the intersection of $A$ and $B$ is connected.}), like $\KHAL{n}/2$ or any hyper-rectangle in $\KHAL{n}/2$, and that it set $Y$ of values is $\KHAL{1}/2$. Then, \emph{threshold sets} are defined for any $\lambda \in Y$ in this way~\cite{boutry2019make,boutry.15.ismm}: 
$$\,\lcut{\U}{\lambda} \, = \, \set{x \in X \,}{\, \forall \, \mu \in \U(x), \; \mu < \lambda}\,$$
$$\,\gcut{\U}{\lambda} \, = \, \set{x \in X \,}{\, \forall \, \mu \in \U(x), \; \mu > \lambda}.$$

We can show (see~\cite{najman.2013.ismm}) that, for any $\lambda \in Y$, the threshold sets $\lcut{\Iu}{\lambda}$ and $\gcut{\Iu}{\lambda}$ are well-composed~\cite{latecki1995well} when we use a \WC interpolation $\Iu$ (see~\cite{boutry.15.ismm,najman.2013.ismm} for examples of \nD \WC interpolations).  Then the combinatorial boundaries of the threshold sets are \emph{discrete surfaces}. By defining respectively the \emph{upper} and the \emph{lower shapes}:
\begin{align*}
&\SHAPESSUP = \left\{\, \sat(\Gamma); \; \Gamma \in \{ \CC(\gcut{\Iu}{\lambda}) \}_\lambda \,\right\},\\
&\SHAPESINF = \left\{\, \sat(\Gamma); \; \Gamma \in \{ \CC(\lcut{\Iu}{\lambda}) \}_\lambda \,\right\},
\label{eq.shapes}
\end{align*}

we obtain then that together these shapes form a set of shapes $\mathfrak{S}_{\mathfrak{I}(u)}$ which is a tree, the \emph{tree of shapes}. 

\subsection{Some elementary properties about shapes}

The following properties of shapes will be useful in the sequel.

\begin{Proposition}[Lemma 2.5 in~\cite{caselles2009geometric}]
\label{propo.sat.increasing}
Let $\dom$ be a topological space. Then, the $\sat$ operator is monotonic, that is, for any subset $A,B$ of $\dom$:
$$A \subseteq B \Rightarrow \sat(A) \subseteq \sat(B).$$
\end{Proposition}

\begin{Proposition}[Lemma 2.4~\cite{caselles2009geometric}]
Let $\dom$ be a topological space. Then, when $A \subseteq \dom$ is an open set, then $\sat(A)$ is open too.
\label{saturation.openset.is.openset}
\end{Proposition}

\begin{Proposition}[Lemma 2.7~\cite{caselles2009geometric}]
Let $\dom$ be a topological space. Then, when $A \subseteq \dom$ is connected, then $\sat(A)$ is connected too.
\label{propo.sat.connected.is.connected}
\end{Proposition}

\begin{Proposition}[Lemma 2.10~\cite{caselles2009geometric}]
\label{propo.sat.inclusion}
Let $\dom$ be a compact, locally connected, and unicoherent topological space. Let $A$ be a subset of $\dom$ such that $\sat(A) \neq \dom$. Then $\sat(A) \subseteq \sat(\partial A)$. Furthermore, if $A$ is closed, $\sat(A) =\sat(\partial A)$.
\end{Proposition}

\begin{figure}
\centering
\includegraphics[width=\linewidth]{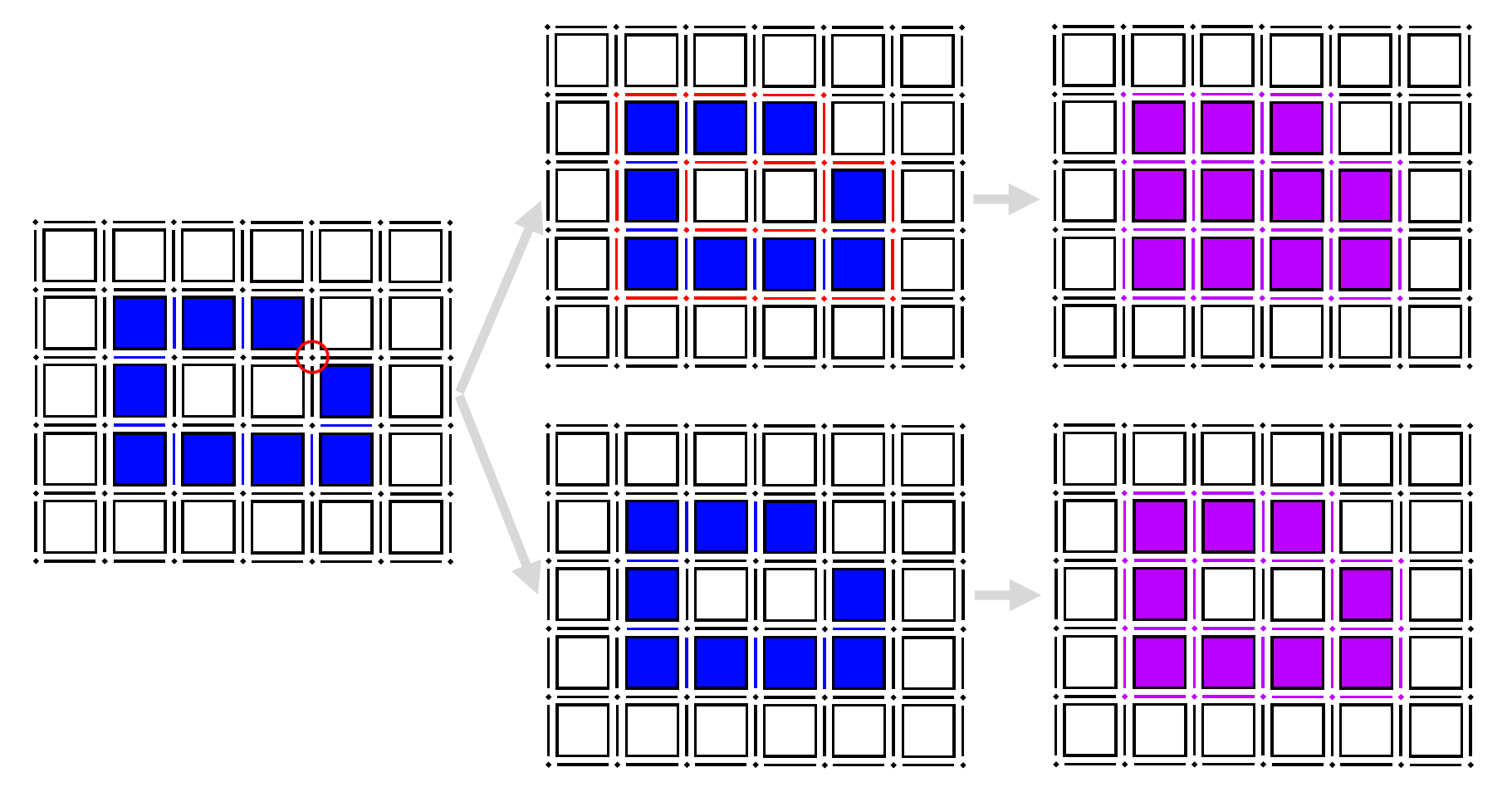}
\caption{The closure and saturation operators do not always commute when the used image is not \WC (see the pinch encircled in red on the left side which causes this topological issue).}
\label{fig.closatKO}
\end{figure}

\begin{figure}
\centering
\includegraphics[width=\linewidth]{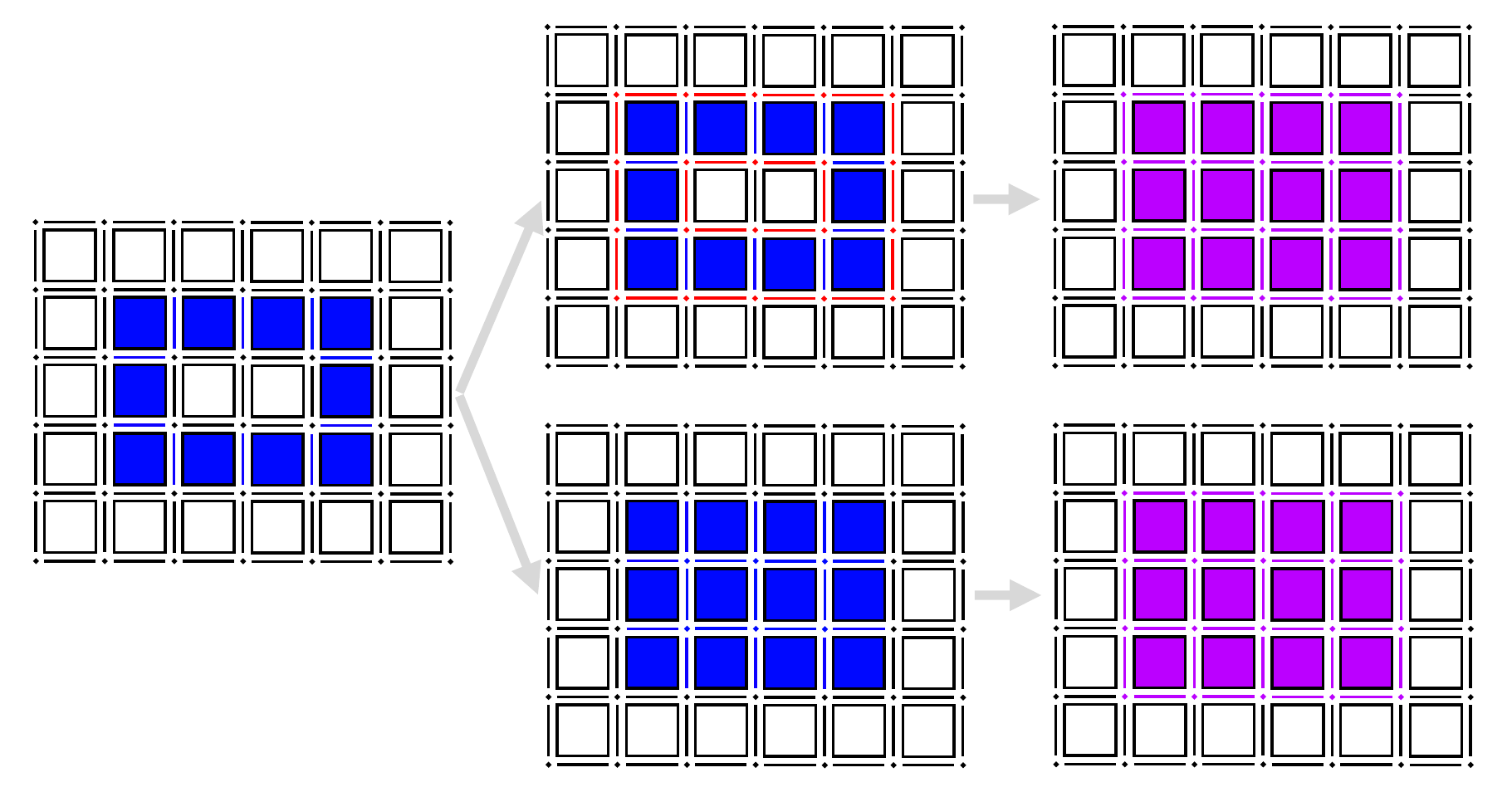}
\caption{The closure and saturation operators commute when the used image is \WC.}
\label{fig.closatOK}
\end{figure}

\newcommand{\clDOM}{cl_{\dom}\xspace}
\newcommand{\stDOM}{st_{\dom}\xspace}

\begin{Proposition}[\cite{najman2013discrete}]
Let $\dom$ be a $n$-D finite (unicoherent) hyper-rectangle in the $n$-D Khalimsky grid, and let $\pinfty$ be a $n$-face of $\dom$ which belongs to $\partial \dom$. When a finite set $X$ is an open regular WC set, then we have the following property for any $x \in \dom$:
$$\sat(\clDOM(X),x) = \clDOM(\sat(X,x)).$$
\label{propo.satCC=CCsat}
\end{Proposition}

Figures~\ref{fig.closatKO} and~\ref{fig.closatOK} depict the possible topological issue which arises when we do not work with \WC images.

\section{Algorithm Description}
\label{sec:algorithm.description}

\subsection{The main algorithm}
\label{ssec:main}

\begin{figure}[htbp]
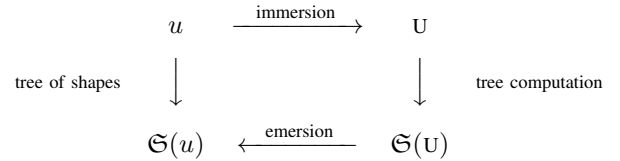

  \centering
  \begin{displaymath}
    \begin{array}{rcccl}
      & u & \xrightarrow{~~\mbox{\scriptsize immersion}~~} & \U & \medskip \\ 
      \mbox{\scriptsize tree of shapes} & \Big\downarrow & & \Big\downarrow & \mbox{\scriptsize tree computation} \medskip \\
      & \Tos(u)  & \xleftarrow{~~\mbox{\scriptsize emersion}~~} & \Tos(\U) &
    \end{array}
  \end{displaymath}
  \caption{Global scheme to compute the tree of shapes.  The ``tree
    computation'' on the right side is composed of the two steps: sort
    (propagation) and union-find (effective tree computation thanks to
    the $\parent$ relationship).  \label{fig.global.scheme} }
\label{fig:summarycomputation}
\end{figure}

Let us begin with an intuitive explanation of the computation of the tree of shapes (see Figure~\ref{fig:summarycomputation}). We start from an image $u$ whose domain is a subset of $\Zeals^n$. Then, we compute in $\Zeals^n/2$ its \WC max-interpolation. Then, we immerse the interpolation into the \nD Khalimsky grid, so we provide some continuity to the new image $\U$ representing $u$. We call abusively this procedure the \emph{immersion step}. Then, based on $\U$, we add a border, from which we start the propagation, and we go deeper and deeper in the image, until we have covered the whole domain of the image. This step is called the \emph{tree computation} because while we cover the domain of the image, we can deduce the parenthood relationship between the components in the image (we prove it in the next section). Since the propagation has been done over a domain which has been subdivided twice (once with the max interpolation and once to immerse $\Zeals^n/2$ into $\KHAL{n}/2$), we have to go back to the initial domain. This last step is called the \emph{emersion} and removes the secondary pixels to keep only the primary pixels.

\subsection{The main procedure of the algorithm more in details}
\label{ssec:sort}

\begin{algorithm}
\caption{The quasi-linear computation of the tree of shapes (main procedure).}
\label{alg.computetree}
\hlinetop
\medskip
\computeqltree($u$) : Pair(T, Array[P]) \;
\Begin{
    \U \setto \immersion(u)\;
  (\Rel, $\ub$) \setto \sort($\U$)\;
  \parent \setto \unionfind(\Rel)\;
  \canonicalizetree(\parent, \Rel, $\ub$)\;
  (\parent', \Rel') \setto \emerge(\parent, \Rel, $\ub$)\;
  \return (\parent', \Rel')\;
} %
\smallskip
\hlinebot
\end{algorithm}

The algorithmic description of the procedure presented in the previous subsection is as described in Alg.~\ref{alg.computetree}. We start with the procedure called \immersion, and we follow with the computation of the tree of shapes in three steps: $(1)$ We use a front propagation algorithm \sort\ which handles a hierarchical queue and starts from the border of the image and covers progressively the whole domain, it outputs then an array \Rel of the ordered pixels of the domain of the image, $(2)$ Now that we have \Rel, we use \unionfind\ to deduce the parenthood relationship \parent between the pixels of the image, that is, it builds the tree of shapes (but without optimization) of $\U$; $(3)$Then the procedure \canonicalizetree\ is utilized to optimize the tree so that each pixel has a parent which is the representative of the component it belongs to. This way, we obtain \parent which is in fact $\Tos(U)$. From it, we can easily deduce $\Tos(u)$ by removing the secondary pixels (see the \emerge\ prodecure).

\subsection{The sorting step}
\label{ssec:sort}

\begin{algorithm}
\caption{The front propagation algorithm sorting the pixels.}
  \label{alg:propagation}
\hlinetop
\medskip

\ppush($q, \, n, \, U, \, \lcur$)\;
\lcomment{modify $q$}\;
\Begin{
  $[\ilower,\,\iupper] \setto U(n)$\;
  \If{$\lcur < \ilower$}{
    $\lcur' \setto \ilower$\;
  }
  \ElseIf{$\lcur > \iupper$}{
    $\lcur' \setto \iupper$\;
  }
  \Else{ 
    $\lcur' \setto \lcur$ \rcomment{we have $\lcur \in U(n)$} \;
  }
  $\push(q[\lcur'], n)$\;
} %
\medskip

\ppop(q, \, \lcur) : H \;
\lcomment{modify $q$, and sometimes $\ell$}\;
\Begin{
  \If{$q[\lcur]$ \textup{is empty}}{
    $\lcur' \setto$ level next to \lcur such as $q[\lcur']$ is not empty\;
    $\lcur \setto \lcur'$\;
  }
  \return $\pop(q[\lcur])$\;
} %
\medskip

\sort($U$) : Pair(Array[H], Image) \;
\Begin{
  \ForAll{h}{
    $\dejavu(h) \setto \false$\;
  }
  $\idx \setto 0$  \rcomment{index in \Rel} \;
  $\linfty \setto U(\pinfty)$ \;
  $\push(q[\linfty],\, \pinfty)$\;
  $\dejavu(\pinfty) \setto \true$\;
  $\lcur \setto \linfty$    \rcomment{start from root level} \;
  \While{$q$ is not empty}{
    $h \setto \ppop(q, \, \lcur)$\;
    $u^\flat(h) \setto \lcur$\;
    $\Rel[\idx] \setto h$\;
    $\idx \setto \idx + 1$\;
    \ForAll{$n \in \mathcal{N}(h) \suchas\ \dejavu(n) = \false$}{
      $\ppush(q,\, n,\, U, \, \lcur)$\;
      $\dejavu(n) \setto \true$\;
    }
  } %
  \return (\Rel, $\ub$)
} %
\smallskip

\hlinebot
\end{algorithm}

The details of the sorting step can be found in Alg~\ref{alg:propagation}. To sort the faces of the domain $X$ of $U$, we use a classical front propagation based on a hierarchical queue~\cite{meyer.1991.afcet}, denoted by $q$, the current level being denoted by $\lcur$. There are two notable differences with the well-known hierarchical-queue-based
propagation.  First the $d$-faces, with $d < n$, are interval-valued
so we have to decide at which (single-valued) level to enqueue those
elements.  The solution is straightforward: a face $h$ is enqueued at
the value of the interval $U(h)$ that is the closest to \lcur (see the
procedure $\ppush$).  Just also note that we memorize the enqueuing
level of faces thanks to the image $u^\flat$ (see the procedure
$\sort$).  Second, when the queue at current level, $q[\lcur]$, is
empty (and when the hierarchical queue $q$ is not yet empty), we shall
decide what the next level to be processed is.  We have the choice of
taking the next level, either less or greater than \lcur, such that
the queue at that level is not empty (see the procedure $\ppop$), even if this choice has no influence on the result \Rel of the algorithm. The image $U$, in addition with the browsing of level in the hierarchical queue, allows for a propagation that is ``continuous'' both in domain space and in level space.

\subsection{Tree Representation}
\label{ssec:tree.representation}

\begin{algorithm}
\caption{Tree construction based on the ancestor relationship \Rel.}
  \label{alg:tree.construction}
\hlinetop
\medskip

\findroot($\zpar, x$) : P
\lcomment{modify \zpar}\;
\Begin{
  \If{\zpar(x) = x}{
    \return $x$
  }
  \Else{
    $\zpar(x) \setto \findroot(\zpar, \zpar(x))$\;
    \return $\zpar(x)$
  }
} %
\medskip

\dounion($p', \, r'$) \;
\lcomment{modify \zpar, \rank, and \last}\;
\Begin{
  \If{$\rank(p') > \rank(r')$}{
    \lcomment{new root is $p'$}\;
    $\zpar(r') \setto p'$\;
    \If{$\last(r') < \last(p')$}{
      $\last(p') \setto \last(r')$
    }
  }
  \Else{
    \lcomment{new root is $r'$}\;
    $\zpar(p') \setto r'$\;
    \If{$\last(p') < \last(r')$}{
      $\last(r') \setto \last(p')$
    }
    \If{$\rank(p') = \rank(r')$}{
      $\rank(r') \setto \rank(r')+1$
    }
  }
} %
\medskip

\unionfind(\Rel) : T \;
\lcomment{with tree balancing and compression}\;
\Begin{
  \ForAll{p}{
    $\zpar(p) \setto \undefTHEO$\;
    $\rank(p) \setto 0$
  }
  \For{$\idx \setto N-1$ \forto 0}{
    $p \setto \Rel[\idx]$  \rcomment{$p$ goes from leaves to root}\;
    $\parent(p) \setto p$\;
    $\zpar(p) \setto p$\;
    $\last(p) \setto \idx$\;
    \ForAll{$n \in \mathcal{N}(p) \suchas\ \zpar(n) \neq \undefTHEO$}{
      $p' \setto \findroot(\zpar, p)$\;
      $r' \setto \findroot(\zpar, n)$\;
      \If{$r' \neq p'$}{
        $r \setto \Rel[\,last(r')\,]$\;
        $\parent(r) \setto p$\;
        $\dounion(p', r')$  \rcomment{update \zpar}\;
      }
    }
  }
  \lcomment{deallocate \zpar, \rank, and \last}\;
  \return \parent
} %
\smallskip

\hlinebot
\end{algorithm}

The max-tree algorithm presented in~\cite{berger.2007.icip} is actually a kind of meta-algorithm
that can be ``filled in'' so that it can serve different aims; in particular, in the present paper, it gives an algorithm to compute the \TOS. An extremely simple union-find structure (attributed by Aho to McIlroy
and Morris) was shown by Tarjan~\cite{tarjan.1975.jacm} to be very
efficient.  This structure, also called disjoint-set data structure or
merge-find set, has many advantages that are detailed
in~\cite{carlinet.2013.ismm}; amongst them, memory compactedness,
simplicity of use, and versatility.  This structure and its related
algorithms are of prime importance to handle connected
operators~\cite{meijster.2002.pami,geraud.2005.ismm}.

\begin{figure*}[h]
\centering
\includegraphics[width=0.65\linewidth]{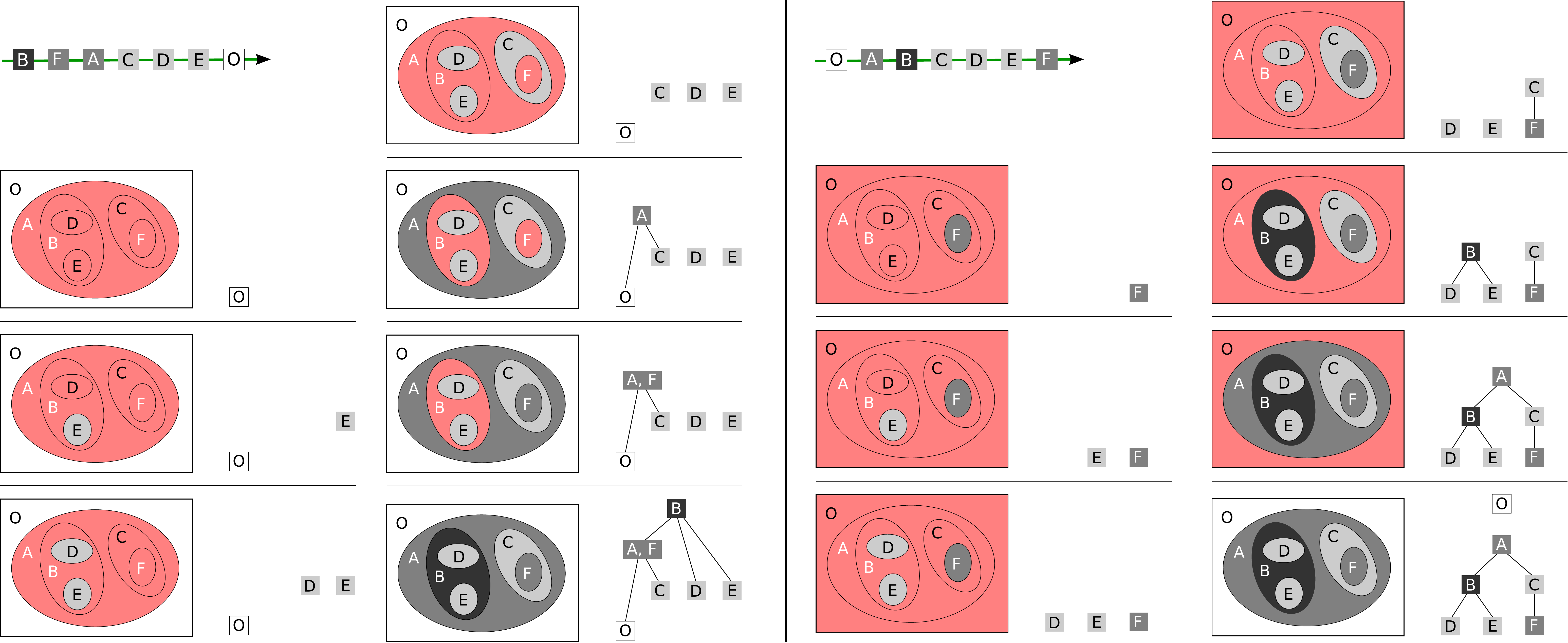}
\caption{Building the max-tree (left side) and the tree of shapes (right side) from root to leaves using the array \Rel (figure extracted from~\cite{geraud.2013.ismm}).}
\label{fig.building}
\end{figure*}

Let us denote by \Rel the ancestor relationship in trees: we have $a
\,\Rel\, p$ iff $a$ is an ancestor of $p$. \Rel can be encoded as an
array of elements (nodes) so that $a \,\Rel\, p \,\Leftrightarrow\,
\indx_\Rel(a) < \indx_\Rel(p)$; browsing that array thus corresponds
to a downwards browsing of the tree, i.e., from root to leaves.  To
construct the max-tree of a given image, we rely on a rooted tree
defined by a parenthood function, named $\parent$, and encoded as an
\nD image (so $\parent(p)$ is an \nD point).  When a node of the
max-tree contains several points, we choose its first point (with
respect to $\Rel$) as the representative for this node; that point is
called a component ``canonical point'' or a ``level root''.  Let
$\Gamma$ denote a component corresponding to a node of the max-tree,
$p_\Gamma$ its canonical element, and $p_r$ the root canonical
element.  The $parent$ function that we want to construct should
verify the following four properties:

\begin{enumerate}
\item %
$\,\parent(p_r) = p_r\,$;
\item %
$\,\forall \, p \neq p_r, \; \parent(p) \, \Rel \, p\,$;
\item %
$\,p$ is a canonical element iff $$\, \{p = p_r\} \, \vee \, \{u(\parent(p)) \neq u(p)\}\,;$$
\item $\,\forall p, p \in \Gamma \Leftrightarrow \{u(p) = u(p_\Gamma)\} \wedge \{\exists
\, i, \parent^i(p) = p_\Gamma\}.$
\end{enumerate}

\begin{algorithm}
\caption{Canonicalization.}
  \label{alg:tree.canonicalization}
\hlinetop
\medskip
\canonicalize($\parent, \Rel$, $\ub$) \;
\lcomment{modify \parent}\;
\Begin{
  \For{$\idx \setto 0$ \forto $N-1$}{
    $p \setto \Rel[\idx]$  \rcomment{$p$ goes from root to leaves}\;
    $p' \setto \parent(p)$\;
    \If{$u^\flat(\parent(p')) = u^\flat(p')$}{
        $\parent(p) \setto \parent(p')$\;
      }
  }
} %
\smallskip
\hlinebot
\end{algorithm}

The routine \textsc{union\_find}, given in
Algorithm~\ref{alg:tree.construction}, is the classical ``union-find''
algorithm~\cite{tarjan.1975.jacm} but \emph{modified} so that it
computes the expected morphological tree~\cite{berger.2007.icip} while
browsing pixels following $\Rel^{-1}$ (see Figure~\ref{fig.building}), i.e., from leaves to root.  Its
result is a $\parent$ function that fulfills those first four
properties. Obtaining the following extra property, ``5.~$\,\forall
p, \; \parent(p)\,$ is a canonical element,'' is extremely interesting
since it ensures that the parent function, when restricted to
canonical elements only, gives a ``compact'' morphological tree.  Precisely it
allows to browse components while discarding their contents: a
traversal is thus limited to one element (one pixel) per component,
instead of passing through every image elements (pixels).
Transforming the parent function so that property~5 is verified can be
performed by a simple post-processing of the union-find computation.
The resulting tree has now the simplest form that we can expect (see Alg.~\ref{alg:tree.canonicalization});
furthermore we have an isomorphism between images and their canonical
representations.

The algorithm presented in~\cite{berger.2007.icip} to compute the max-tree is also able to compute the \TOS. The skeleton of this algorithm is the routine \textsc{compute\_tree} is composed of the steps presented above. In the case of the max-tree, the sorting step provides \Rel encoded as
an array of points sorted by increasing gray-levels in $u$, i.e., such
that the array indices satisfy $~\idx < \idx' \;\Rightarrow\; u(\Rel[\idx]) \leq u(\Rel[\idx'])$. Last, the canonicalization post-processing is trivial~\cite{berger.2007.icip}. In the case of the \TOS, it is also a tree that represents an inclusion relationship between connected
components of the input image.  As a consequence an important
idea to catch is that the \TOS can be computed with the exact same
routine, \textsc{union\_find}, as the one used by the max-tree. The major and crucial difference between the max-tree and the \TOS
computations is obviously the sorting step (see Figure~\ref{fig.building}). For the \textsc{union\_find} routine to be able to compute the \TOS using
$\Rel^{-1}$, the \textsc{sort} routine has to sort the image elements
so that \Rel corresponds to a downward browsing of the \TOS.
Schematically, \Rel contains the image pixels going from
the ``external'' shapes to the ``internal'' ones (included in the former ones).

\subsection{The last step: the emersion}

\begin{algorithm}
\caption{Emersion procedure used to obtain the final tree (in the domain of the initial image).}
  \label{alg:tree.emersion}
\hlinetop
\medskip

\isrepresentative($x$) : $\mathbb{B}$ \;
\Begin{
  \return $\parent(x) = x ~\myor u^\flat(\parent(x)) \neq u^\flat(x)$
} %
\medskip

\getrepresentative($x$) : P \;
\Begin{
  \While{$\mynot \isrepresentative(x)$}{
    $x \setto \parent(x)$\;
  }
  \return $x$
} %
\medskip

\findprimaryancestor($x$) : P \;
\Begin{
  \Repeat{$\mynot \isprimary(x) ~\myand \parent(x) \neq x$}{
    $x \setto \parent(x)$\;
  }
  \return $x$
} %
\medskip

\emerge(\parent, \Rel, $\ub$)\;
\Begin{
  \lcomment{rearrange}\;
  \For{$i \setto 0$ \forto $N-1$}{
    $p \setto \Rel[i]$  \rcomment{$p$ goes from root to leaves}\;
    \If{$\isrepresentative(p)$}{
      \continue \;
    }
    $p' \setto \getrepresentative(p)$\;
    \If{$\mynot \isprimary(p') ~\myand \isprimary(p)$}{
      \If{$\parent(p') = p'$}{
        $\parent(p) \setto p$\;
      }
      \Else{
        $\parent(p) \setto \parent(p')$\;
      }
      $\parent(p') \setto p$\;
    }
  }

  \lcomment{keep only primary points}\;
  $j \setto 0$\;
  \For{$i \setto 0$ \forto $N-1$}{
    $p \setto \Rel[i]$  \rcomment{$p$ goes from root to leaves}\;
    \If{$\isprimary(p)$}{
      $\Rel[j] \setto p$\;
      $j \setto j + 1$\;
      $\parent(p) \setto \findprimaryancestor(p)$\;
    }
  }
  $N \setto j$\;

  \lcomment{canonicalization}\;
  \canonicalize($\parent, \Rel$, $\ub$)\;

  \return (\parent, \Rel)
} %
\smallskip

\hlinebot
\end{algorithm}

Using the canonicalized tree, we can finalize the procedure and emerge the connected components of the tree simply by removing the secondary pixels of the components (see Algo~\ref{alg:tree.emersion}).

\section{The $n$-D proof}
\subsection{New mathematical properties}

Let us recall that the strict threshold sets of a plain map are open sets.

\begin{figure}[htb!]
\begin{center}
\includegraphics[width=0.88\linewidth]{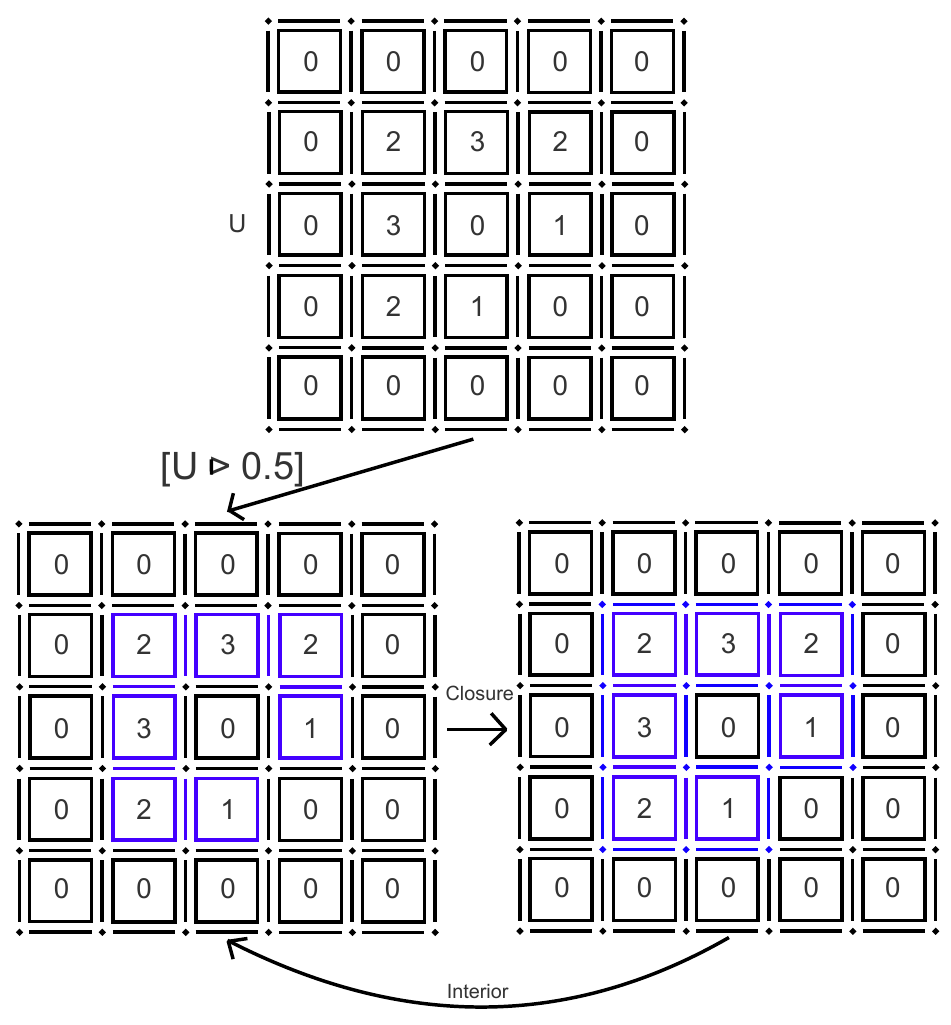}
\end{center}
\caption{Any strict threshold set of a \SBIPMAP is regular}
\label{fig:regular}
\end{figure}

\begin{Proposition}
Let $U$ be some \SBIPMAP on a $n$-D Khalimsky grid, then any of its strict threshold sets $T$ is regular, that is, it satisfies:
$$\Interior(\Cl{T}) = T.$$
The direct consequence is that its strict shapes are regular sets too.
\label{propo.plain.maps.regular}
\end{Proposition}

\IEEEproof{Let us treat the upper case (see Figure~\ref{fig:regular}) for some value $\lambda \in \Reals$ (the lower case follows the same reasoning). Let $\lambda$ be some real value, and $[U \rhd \lambda]$ the corresponding lower set. Now, let us prove its regularity using a double inclusion. First, for any $\lambda \in \Reals$, $[U \rhd \lambda]$ is contained in $\Cl{[U \rhd \lambda]}$, thus $[U \rhd \lambda] = \Interior([U \rhd \lambda]) \subseteq \Interior(\Cl{[U \rhd \lambda]})$, which proves the first inclusion. Second, let us define some $z \in \Interior(\Cl{[u \rhd \lambda})$, then $\St{z} \subseteq \Cl{[U\rhd\lambda]}$. The subset $(\St{z})_n$ made of the $n$-faces of $\St{z}$ is contained in $\Cl{[U \rhd \lambda]} = \partial[U \rhd \lambda] \sqcup [U \rhd \lambda]$ (disjoint union). Since $\partial [U \rhd \lambda]$ does not contain any $n$-face (as every combinatorial boundary), then:
$$(\St{z})_n \subseteq [U \rhd \lambda].$$
Consequently, for any $p \in (\St{z})_n$, $U(p) \rhd \lambda$, which leads by using the formula of the span-based immersion to:
$$U(z) := \SPAN\{U(p) \; ; \; p \in (\St{z})_n\} \rhd \lambda,$$
then $U(z) \rhd \lambda$, that is, $z \in [U \rhd \lambda]$. The proof is done.}

\begin{figure}[htb!]
\begin{center}
\includegraphics[width=0.88\linewidth]{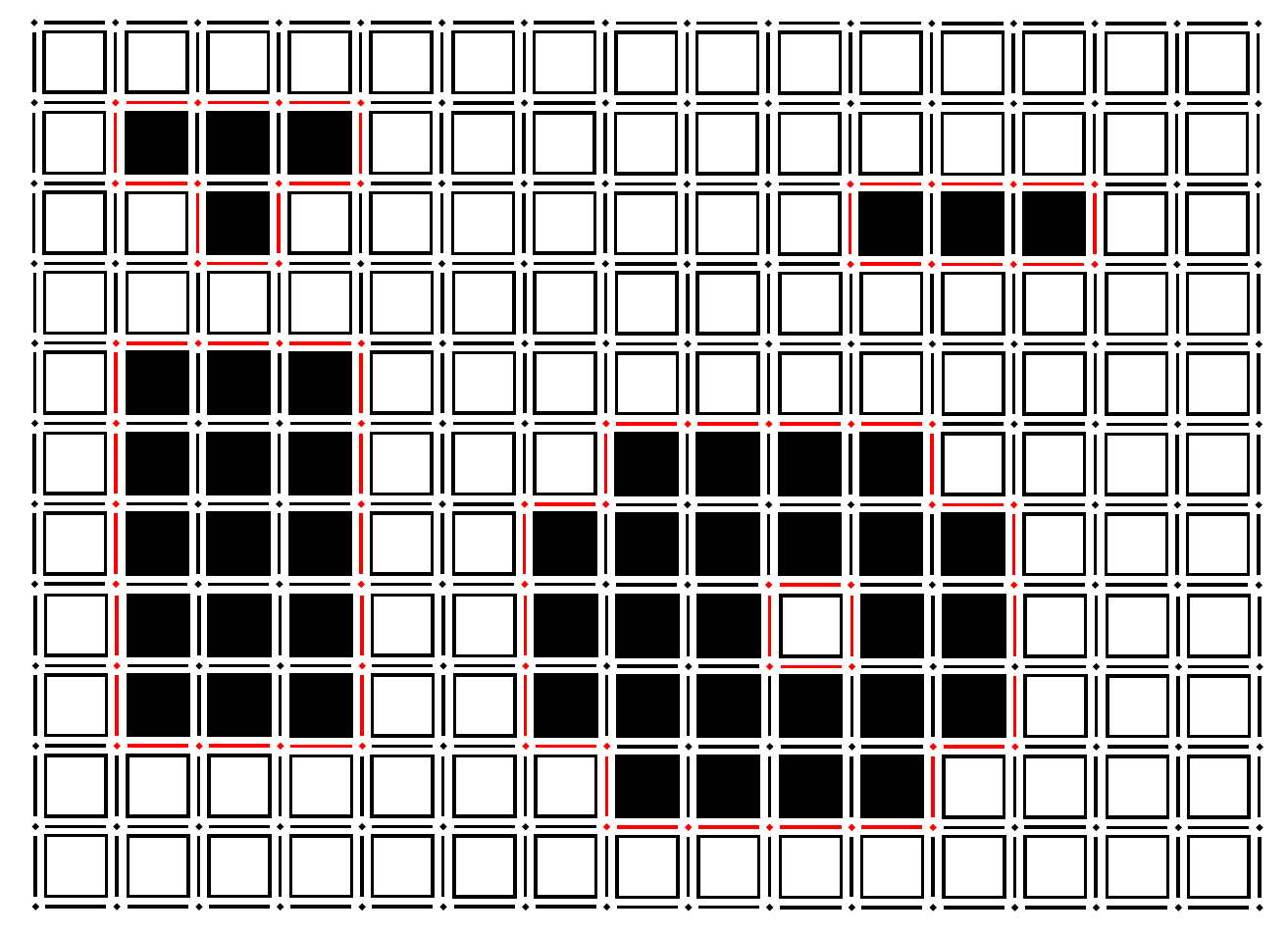}
\end{center}
\caption{Boundaries in red of some subsets of $\KHAL{2}$.}
\label{fig:boundaries}
\end{figure}

  \begin{figure}[htb!]
    \begin{center}
      \includegraphics[width=0.88\linewidth]{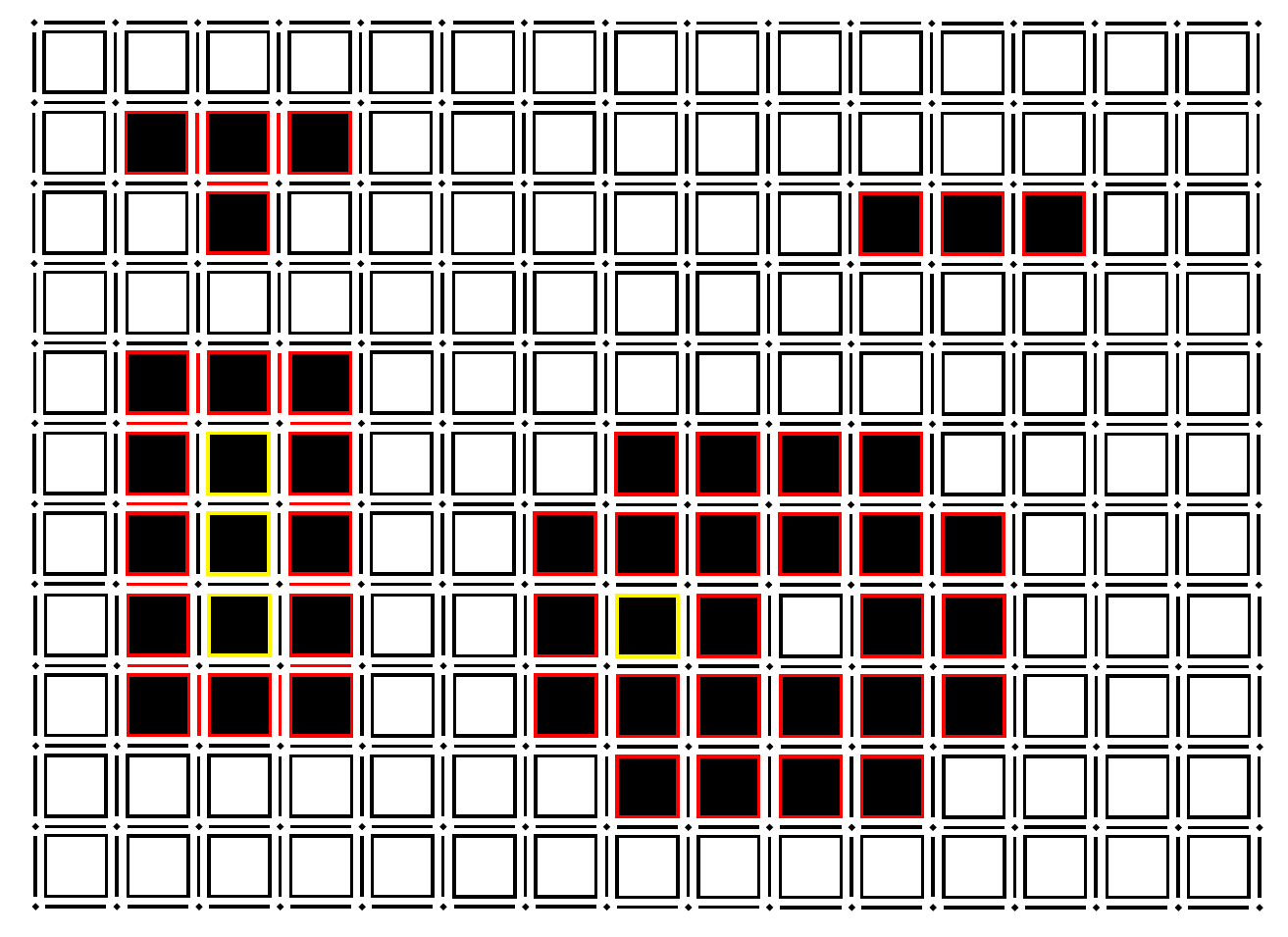}
      \caption{Internal boundaries (in black square encircled in red) of some subsets of $\KHAL{2}$; the $n$-faces remaining after having removed the internal boundaries to their respective sets are encircled in yellow to show the thickness of the internal boundaries.}
    \end{center}
\label{fig:boundaries}
  \end{figure}

\newcommand{\clHN}{cl_{\KHAL{n}}\xspace}
\newcommand{\stHN}{st_{\KHAL{n}}\xspace}

\begin{Definition}[Interior boundaries]
Let $\KHAL{n}$ be the $n$-D Khalimsky grid, and let $F$ be an open subset of $\KHAL{n}$. We call \emph{interior boundary} of $F$ the set $\intbd F = \stHN(\bd F) \cap F$.
\end{Definition}

\newcommand{\clX}{cl_X\xspace}
\newcommand{\stX}{st_X\xspace}

We recall that a family $\bigcup_{i \in I} X_i$ of subsets of a poset $X$ is said to be a separated union when for any $i,j$ in $I$ with $i \neq j$, $(\stX(X_i) \cup \clX(X_i)) \cap X_j = \emptyset$.

\subsection{Property of the internal boundaries}

\begin{Proposition}
Let $\cav$ be an open subset of $\KHAL{n}$. The set $\intbd(\cav)$ can be reformulated this way:
$$\intbd(\cav) = \stHN(\cav) \cap \stHN(\KHAL{n} \setminus \cav).$$
\label{proposition.reformulation.interior.boundary}
\end{Proposition}

\IEEEproof{Let us prove this equality by a double inclusion relationship. When $h$ is an element of $\intbd (\cav) = \st(\partial \cav) \cap \cav$, then $h$ belongs to $\cav \subseteq \st(\cav)$ and then $h \in \st(\cav)$. In addition, $\KHAL{n} \setminus \cav$ is closed since $\cav$ is open by hypothesis, which means that $\KHAL{n} \setminus \cav$ contains its boundary $\partial \cav$. This way, we have that $h \in \st(\partial \cav) \subseteq \st(\KHAL{n} \setminus \cav)$. This concludes the first implication. Conversely, when $h$ belongs to $\st(\cav) \cap \st(\KHAL{n} \setminus \cav)$, then $h$ belongs to $\st(\cav) = \cav$. Furthermore, $h$ belongs to $\st(\KHAL{n} \setminus \cav)$ and then contains (as a face) some other face of $\KHAL{n} \setminus \cav$. However, the only elements of $\KHAL{n} \setminus \cav$ which are adjacent to $\cav$ are the elements of $\partial T$, then $h$ contains a face $z$ of $\partial \cav$, then $h \in \st(\partial \cav)$. This proves the second implication.}

\subsection{Domains of the \SBIPMAPS and their (thick) border}
\label{sec.border}

\newcommand{\UEXT}{U_{ext}\xspace}

From now on, we will use the following notations.

\medskip

Let the domain $\dom$ of the given \SBIPMAP $U$ be some non-empty bounded open hyper-rectangular domain in the $n$-D Khalimsky grid $\KHAL{n}$.

Then, we deduce the \textquote{enlarged} domain:
$$\CLOSEDDOM := \clHN(\stHN(\clHN(\dom))),$$
whose \emph{border} $\border \CLOSEDDOM$ is the set:
$$\border \CLOSEDDOM = \CLOSEDDOM \setminus \dom.$$

We naturally extend $U$ to $\UEXT$ on $\CLOSEDDOM$ in the following way: we set at $\{\linfty\}$ with $\linfty \in \Reals$ all the $n$-faces of the border of $\CLOSEDDOM$ and we choose arbitrarily one of the $n$-faces of $\border \CLOSEDDOM$ as exterior point $\pinfty$. The faces of $\CLOSEDDOM \setminus \KHAL{n}_n$ are then naturally set using a procedure detailed later and which ensures the continuity of the new set-valued image $\UEXT$ defined on $\CLOSEDDOM$.

This way, we ensure that any shape $\shape$ of $\UEXT$ is either the whole domain $\CLOSEDDOM$ or a subset of $\dom$. We are thus ensured that for any shape $\shape$ of $\UEXT$, the boundary $\partial \shape$ is equal to $\partial \CLOSEDDOM$ or does not intersect $\partial \CLOSEDDOM$. Furthermore, the complementary in $\CLOSEDDOM$ of any shape $\shape$ of any \SBIPMAP defined on $\CLOSEDDOM$ is either empty or connected.

For sake of simplicity, we will refer to $\UEXT$ by calling $U$.

\newcommand{\PR}{\ensuremath{\mathcal{PR}}\xspace}
\newcommand{\PCAV}{\PR-cavity\xspace}
\newcommand{\PCAVS}{\PR-cavities\xspace}

\subsection{\PCAVS and their mathematical properties}

\begin{figure}[htb!]
    \centering
    \includegraphics[width=0.7\linewidth]{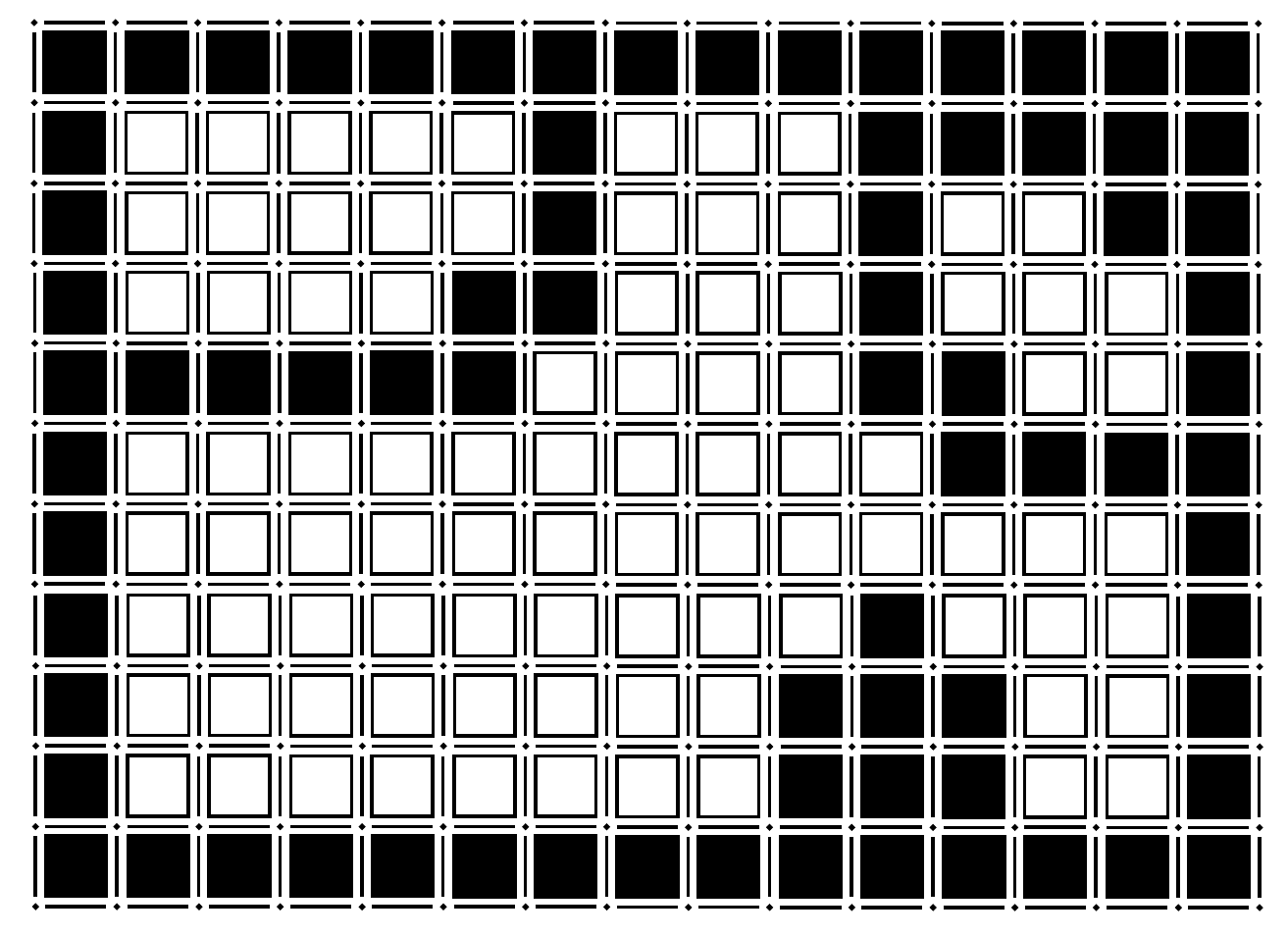}
    \caption{Example of $\front_0$ and its three cavities}
    \label{example.P0}
\end{figure}

\begin{figure}[htb!]
    \centering
    \includegraphics[width=0.7\linewidth]{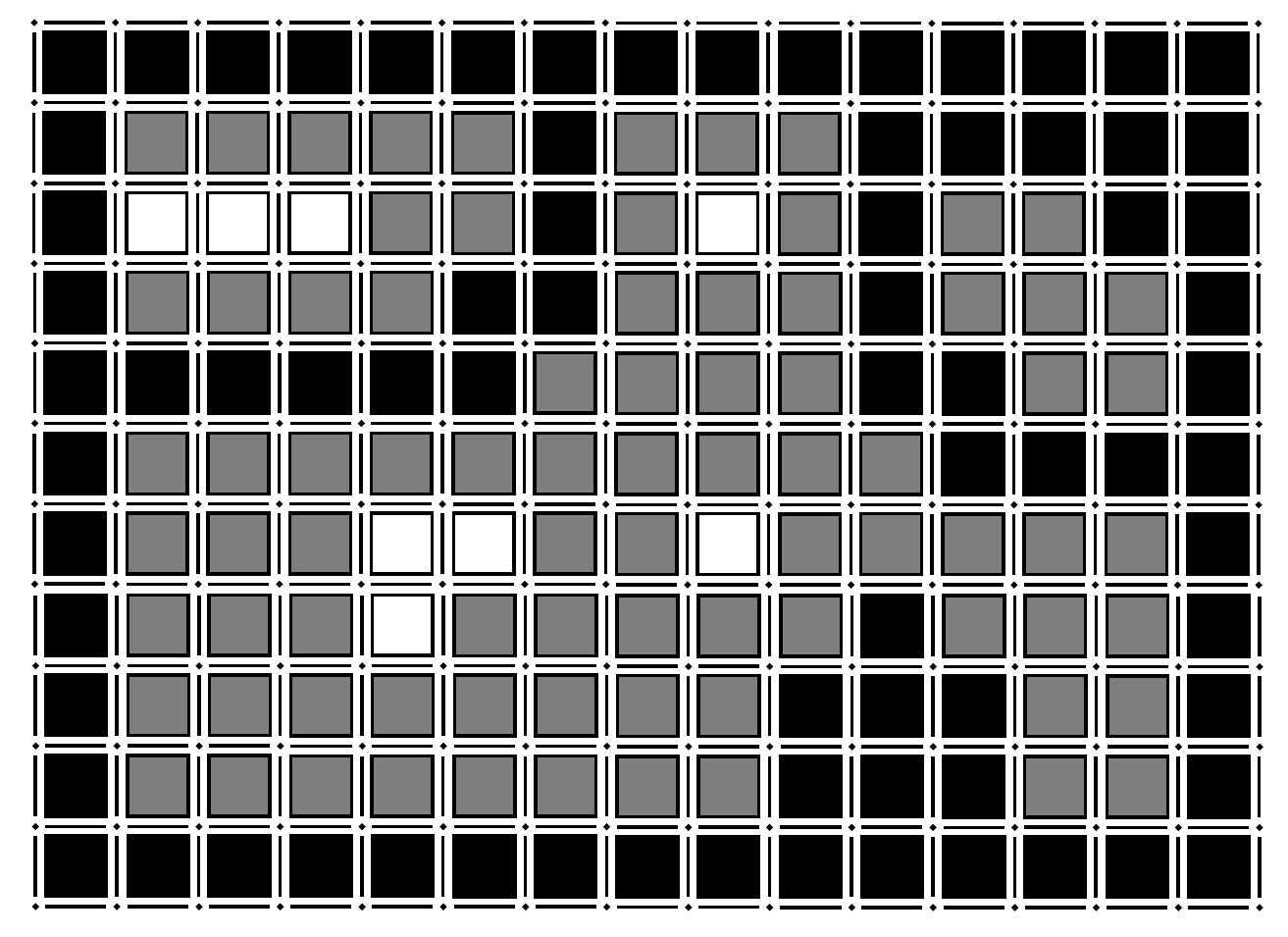}
    \caption{Example of $\front_1$ and its cavities: compared to $\front_0$, some cavities appeared and some disappeared.}
    \label{example.P1}
\end{figure}

\begin{figure}[htb!]
    \centering
    \includegraphics[width=0.7\linewidth]{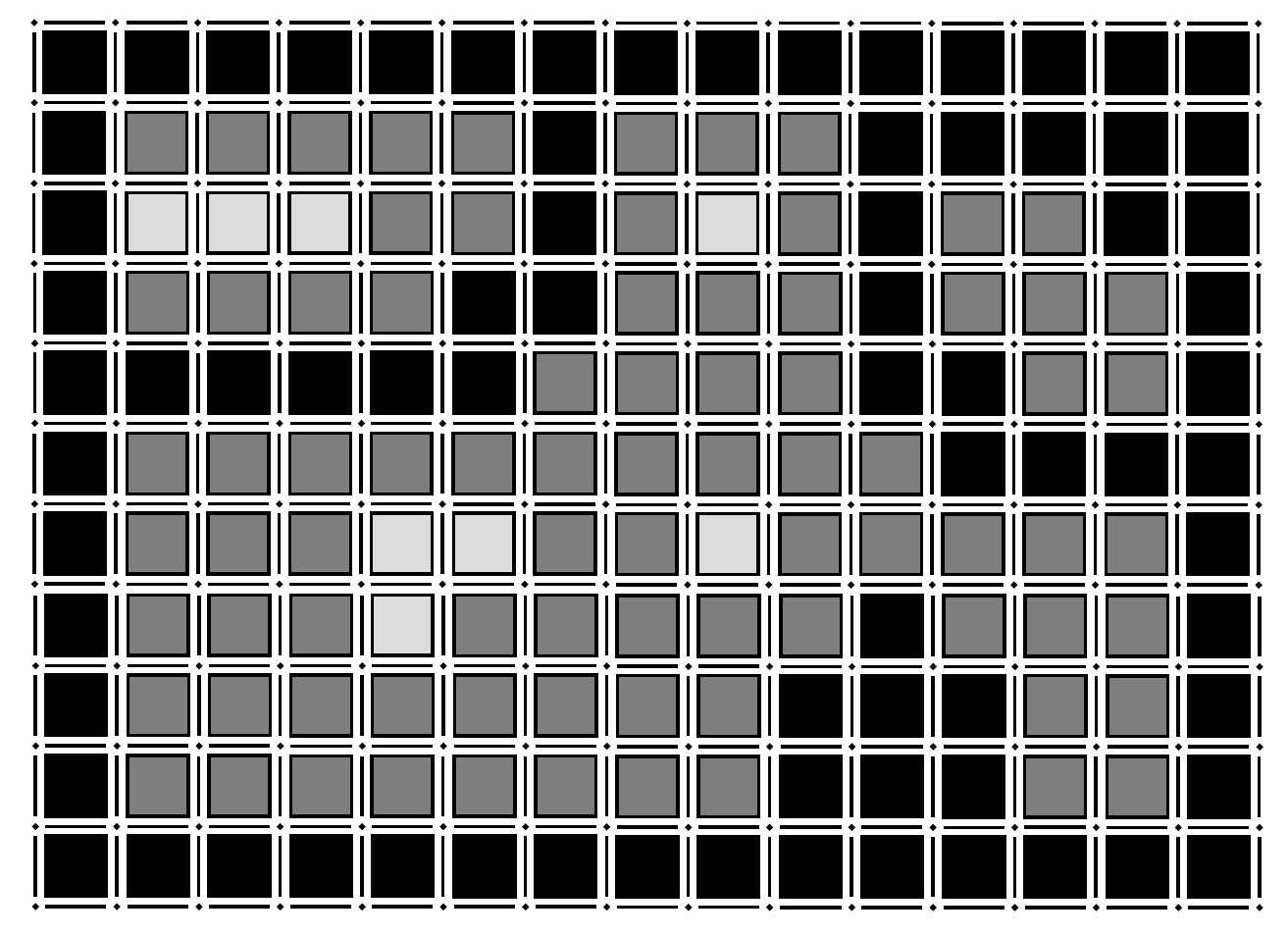}
    \caption{Example of $\front_2$ where the whole domain has been covered.}
    \label{example.P2}
\end{figure}

\begin{Notation} Assuming that $\level_i$ is the level at step $i$ of the front propagation algorithm (FPA), we can reformulate the front propagation (see Figures~\ref{example.P0},\ref{example.P1} and~\ref{example.P2}) this way:

\begin{itemize}

    \setlength\itemsep{1em}

    \item $\leveli{0} = \linfty$ is the level at step $0$,
    
    \item $\front_{0} = \CC([U = \linfty],\pinfty)$ is the covered part of $\CLOSEDDOM$ at step $0$,
    
    \item $\{\cavij{0}{j}\}_{j \in \JSET(0)} = \CC(\CLOSEDDOM \setminus \front_0)$ is the set of what we call \emph{\PCAVS} at step $0$,

    \item when $i > 0$, $$
    \front_{i+1} = \front_{i} \cup \bigcup_{j \in \JSET(i)}\bigcup_{x \in \intbd(\cavij{i}{j})} \CC([U = \leveli{i+1}],x)
    $$
    
    is the covered domain at step $i$,

    \item when $i > 0$,
    $$\{\cavij{i+1}{j}\}_{j \in \JSET(i+1)} = \CC(\CLOSEDDOM \setminus \front_{i+1})$$
    is the set of what we call \emph{\PCAVS} at step $i > 0$.

\end{itemize}
\end{Notation}

\underline{Note:} the elements $\front_i$ are directly deduced from the propagation presented in Alg.~\ref{alg:propagation} and the terms $\cavij{i}{j}$ corresponds to cavities in $\KHAL{n}$ but will be called \PCAVS to emphasize that they are obtained thanks to the propagation algorithm.

\medskip

\begin{figure}[!ht]
    \centering
    \includegraphics[width=0.7\linewidth]{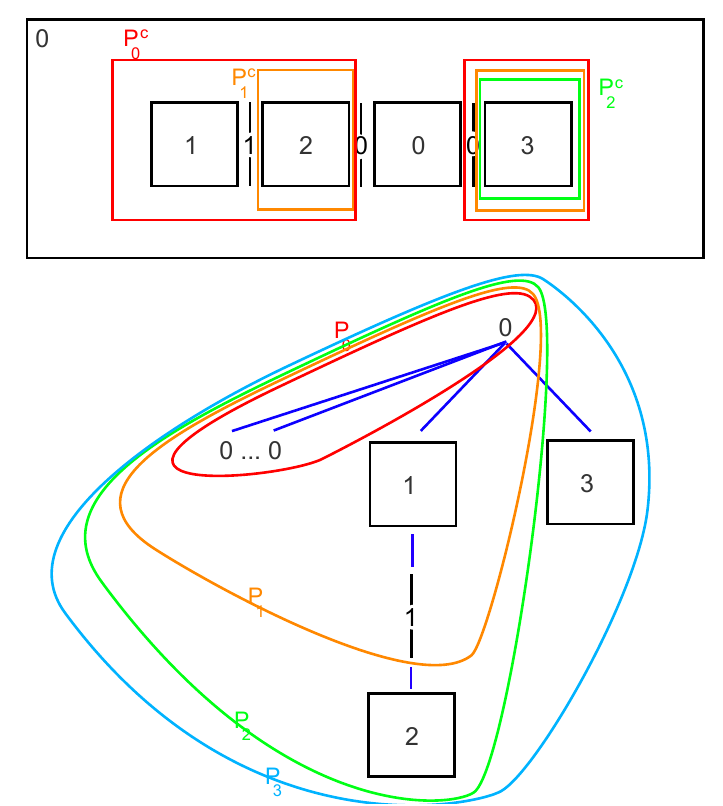}
    \caption{Two equivalent representations of the same data on a toy example: at the top, the image $\ub$ in a 2D Khalimsky grid, and at the bottom, the corresponding $\PARENTSTRUCTURE$ structure (in dark blue) resulting from the union-find algorithm; the corresponding output from the sort algorithm is $\mathcal{R} = [0,\dots,0,1,1,2,3]$, and the levels $\level$ are in the following order: $[0,1,2,3]$.  While $\front_i$ is progressively covering the whole domain $\CLOSEDDOM$, the structure $\PARENTSTRUCTURE$ grows and represents the way the propagation covers the domain. When a node in $\PARENTSTRUCTURE$ has a value different from the one of its parent, a \emph{level-line} is crossed, which means that the depth of this branch into the image increases.}
    \label{fig.2representations}
\end{figure}

Now, let us briefly show that the representation $\PARENTSTRUCTURE$ provided by the union-find algorithm and the image $\ub$ provided by the front propagation algorithm are equivalent. They both encode the front propagation, as depicted in Figure~\ref{fig.2representations}.

\medskip

Furthermore, they show the following properties:

\begin{Property}
At each step $i$ of the front propagation algorithm, the front $\front_i$ is connected.
\label{property.parentUF.1}
\end{Property}

\IEEEproof{First, the tree resulting from the propagation is connected by construction. Second, two nodes are connected in this tree when they satisfy a covering relationship in the Khalimsky grid (due to the propagation based on the $2n$-connectivity). Thus, the representation of the tree in the domain $\CLOSEDDOM$ is connected, since a covering relation implies adjacency.}

\begin{Property}
At each step $i$ of the front propagation algorithm, the front $\front_i$ is closed.
\label{property.parentUF.2}
\end{Property}

\IEEEproof{This results from the fact that for any face $h \in \CLOSEDDOM$ which is treated by the front propagation algorithm, all its faces $h'$ will satisfy $\ub(h) \in U(h')$, thus all the faces $h'$ will be treated at the same step (with the same current level $\level$). In other words, when $h$ belongs to $\front_i$, all the closure of $\front_i$ is contained in $\front_i$, which leads to the fact that $\front_i$ is closed.}

\begin{Property}[\PCAVS form a hierarchy]
\label{property.cav.tree}
The set of \PCAVS of a \SBIPMAP $U$ defined on a domain $\CLOSEDDOM$ satisfies that for two \PCAVS $\cav,\cav'$ such that $\cav \cap \cav' \neq \emptyset$, then we have $\cav \subseteq \cav'$ or $\cav \supseteq \cav'$. In other words, the set of \PCAVS of $U$ is a tree, and we will denote it by $\HierarchyCavities$.
\end{Property}

\IEEEproof{By increasingness of the front $\front_i$, we obtain easily that the connected components of its complementary are nested. Furthermore, $\CLOSEDDOM$ is a superset of any other cavity of $U$. Thus, $\HierarchyCavities$ is a tree.}

\medskip

\begin{Notation}
Let $\cav \subsetneq \CLOSEDDOM$ be a cavity of a \SBIPMAP $U$. We denote by $\parentCAV(\cav)$ the direct parent of $\cav$ in $\HierarchyCavities$. The root of $\HierarchyCavities$, $\CLOSEDDOM$, has no parent in $\HierarchyCavities$.
\end{Notation}

\begin{Notation}
Let $\cav$ be a \PCAV of a \SBIPMAP $U$. We define the following set: $\nextCAV(\cav) = $
$$\left\{\cav' \in \HierarchyCavities(U) \; ; \; \cav = \parentCAV(\cav')\right\}.$$
\end{Notation}

\begin{Notation}
Let $\cav$ be a \PCAV of a \SBIPMAP $U$. We denote by $\childrenCAVITY(\cav)$ the set of children of $\cav$ in $\HierarchyCavities(U)$.
\end{Notation}

\begin{Proposition}
Let $U$ be some \SBIPMAP on $\CLOSEDDOM$. For any \PCAV $\cav$ of $U$, we have the following equality:
$$\bigcup_{\cav' \in \nextCAV(\cav)} \cav'  = \bigcup_{\cav' \in \childrenCAVITY(\cav)} \cav'.$$
\label{propo.cav}
\end{Proposition}

\IEEEproof{The first inclusion is immediate. The converse inclusion follows directly from Property~\ref{property.cav.tree}.}

\begin{figure}[htb!]
\begin{center}
\includegraphics[width=0.88\linewidth]{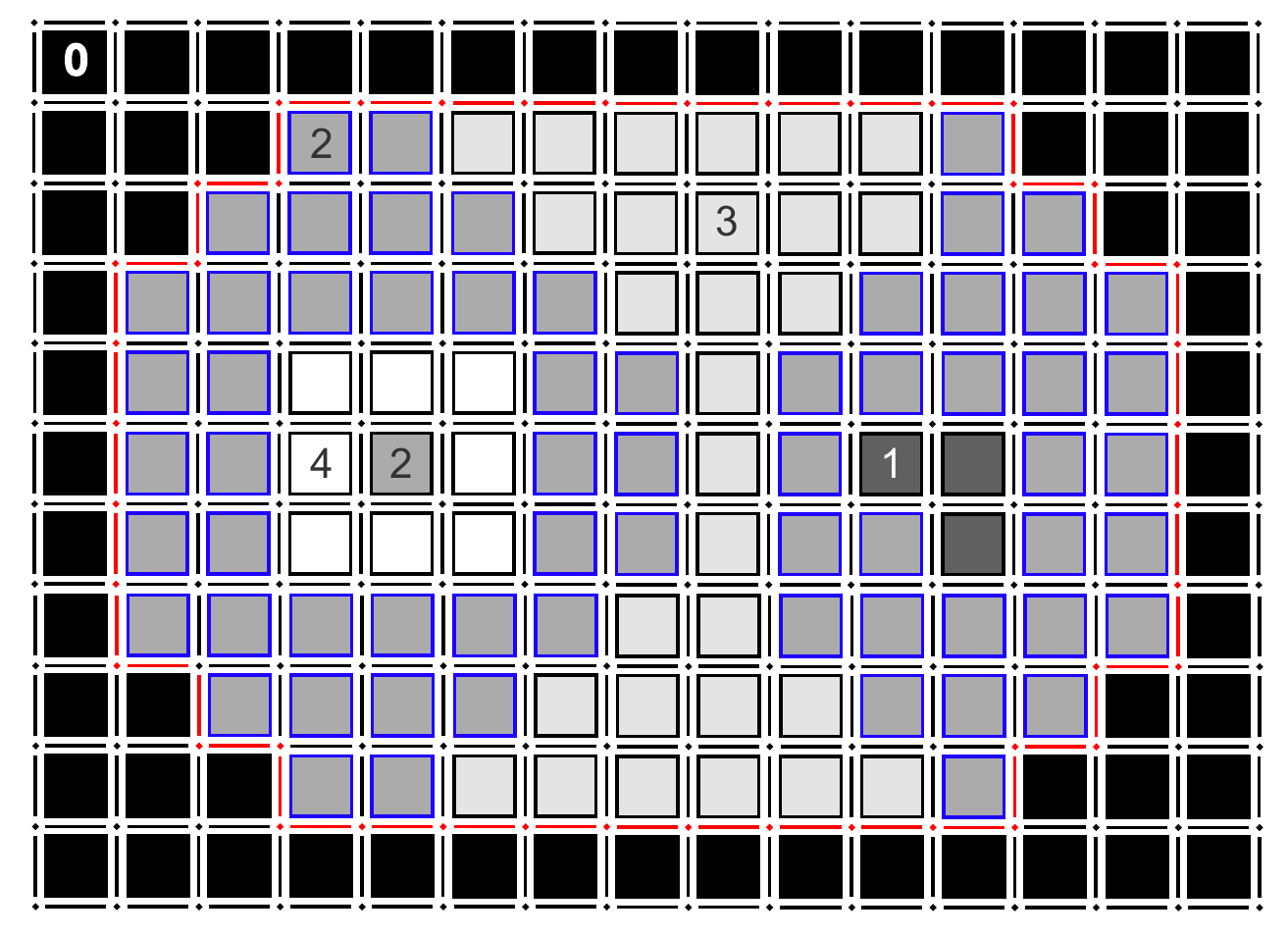}
\end{center}
\caption{The front $\front_0$ is depicted in black and shows the propagation on the domain of a \SBIPMAP $U$ at the initial step. Then, the remaining (open) \PCAV $\cav$ (whose boundary is depicted in red) will be partially covered at the next propagation level $\ell = 2$ since it is the nearest level among $\{2,3\}$. We can thus affirm that $\levelCAV(\cav) = 2$. Furthermore, the set $\properCAV{\cav}$ is depicted by all the blue faces connected into $[U = 2]$ to the pixels of value $2$ in the internal boundary of $\cav$. This set is also equal to $\cav$ minus its children in the hierarchy constructed thanks to the propagation.}
\label{fig:own_elements_cav}
\end{figure}

The following definition is depicted in Figure~\ref{fig:own_elements_cav}.

\begin{Definition}
Let $U$ be some \SBIPMAP on $\CLOSEDDOM$. For some \PCAV $\cav$ of $U$, we will denote by $\levelCAV(\cav)$ the level at which the FPA covered $\cav$ starting from the elements of its internal boundary and we call it the \emph{proper level} of $\cav$.
\end{Definition}

\begin{Property}
Any \PCAV $\cav$ of a \SBIPMAP $U$ is an open subset of $\KHAL{n}$.
\label{property.holes.are.open}
\end{Property}

\IEEEproof{Let $i$ be some fixed value in $[0,N]$. Since $P_i$ is built from a finite union of closed sets, it is closed. The direct consequence is that its complementary is an open set, and then each of its connected components is an open set too.}

\begin{Property}
The internal boundary of each \PCAV $\cav$ of a \SBIPMAP $U$ is connected.
\label{property.interiorboundary.connected}
\end{Property}

\IEEEproof{Let us notice that $\intbd \cavij{i}{j}$ is the intersection of $$P_i \cup \bigcup_{k \in \JSET(i)\setminus\{j\}} \cavij{i}{k} \cup \intbd \cavij{i}{j}$$ (which is an open connected set) and of $\cavij{i}{j}$ (which is open by Property~\ref{property.holes.are.open} and connected by definition). Furthermore, their intersection is equal to $\intbd \cavij{i}{j}$, and their union covers $\CLOSEDDOM$. Since $\CLOSEDDOM$ is unicoherent by hypothesis, then $\intbd \cavij{i}{j}$ is connected.}

\begin{Definition}
Let $U$ be a \SBIPMAP on $\CLOSEDDOM$. We say that a \PCAV $\cav$ different from $\CLOSEDDOM$ is an \emph{upper \PCAV} when the proper level of $\cav$ is greater than the one of its parent (positive polarity). At the opposite, we say $\cav$ is a \emph{lower \PCAV} when its proper level is lower than the one of its parent (negative polarity). Two \PCAVS are said to be \emph{of the same nature} when they are both upper \PCAVS or both lower \PCAVS.
\end{Definition}

\begin{Proposition}
Let $U$ be a \SBIPMAP on $\CLOSEDDOM$ and $\cav$ be a \PCAV of $U$ different from $\CLOSEDDOM$. Let $h$ be an element of $\partial \cav$ and let $\parentCAV^k(\cav)$ with $k \geq 1$ be the nearest parent of $\cav$ containing $h$. Then,
$$(\cav, \parentCAV(\cav),\dots,\parentCAV^{k-1}(\cav))$$ have the same nature (the proper levels are monotonic relatively to $k$). The direct consequence is that $U(h)$ contains:
$$[\levelCAV(\parentCAV^{k}(\cav)),\levelCAV(\cav)]$$ in the upper case, and contains:
$$[\levelCAV(\cav),\levelCAV(\parentCAV^{k}(\cav))]$$ in the lower case.
\label{propo.polarity.cte}
\end{Proposition}

\IEEEproof{Let $\cav$ be some \PCAV different from $\CLOSEDDOM$, and let $h$ be one face of $\partial \cav$. We assume without constraint that the first parent of $\cav$ containing $h$, denoted by $\parentCAV^k(\cav)$, is $\CLOSEDDOM$. Let us treat the case where $\cav$ has a positive polarity (its proper level is greater than the one of its parent). We want to prove that the sequence $\left( \levelCAV(\parentCAV^l(\cav)) \right)_{l \in [1,k-1]}$ is strictly decreasing. 

\medskip

For this aim, let us assume that:
$$\min_{l \in [1,k-1]} \levelCAV(\parentCAV^l(\cav)) \leq \linfty,$$
then we can observe easily that the level line corresponding to $\linfty$ will separate $\cav$ and one of its parents, that is, they will be in separate branches in the hierarchy $\HierarchyCavities$, which is impossible by hypothesis.

\medskip

We can now affirm that $\min_{l \in [1,k-1]} \levelCAV(\parentCAV^l(\cav)) > \linfty$. Now, we want to prove that the sequence $\left( \levelCAV(\parentCAV^l(\cav)) \right)_{l \in [1,k-1]}$ is strictly decreasing. For this aim, we assume that it is not true and thus we choose the parent $\parentCAV^l(\cav)$ of $\cav$ in this sequence whose polarity is negative and whose level $\levelCAV$ is minimal. We have then:
$$\levelCAV(\parentCAV^{l-1}(\cav)) > \levelCAV(\parentCAV^l(\cav)).$$
It is clear that during the propagation, when the current level increases to cover $\cav$, it will reach the level line corresponding to $\levelCAV(\parentCAV^l(\cav))$ before reaching $\levelCAV(\parentCAV^{l-1}(\cav))$ and then we will not obtain that $\parentCAV^l(\cav)$ is the parent of $\parentCAV^{l-1}(\cav)$, which is a contradiction. The proof is done.}

\begin{Proposition}
\label{propo.holes.have.uniform.contour.lbl}
Let $\cavij{i}{j}$ be a \PCAV of $U$ with $i \geq 0$ and $j \in \JSET(i)$. We call $\lpar = \levelCAV(\parentCAV(\cavij{i}{j}))$ the level of its parent as a \PCAV of $U$. Then, either $\levelij{i}{j} > \lpar$, and we obtain that:
$$\forall x \in \intbd(\cavij{i}{j}), U(x) \rhd \lpar,$$
or $\levelij{i}{j} < \lpar$, and we obtain that:
$$\forall x \in \intbd(\cavij{i}{j}), U(x) \lhd \lpar.$$
In both cases, we have:
$$\forall x \in \bd\cavij{i}{j}, \lpar \in U(x).$$
\end{Proposition}

\IEEEproof{Let us proceed by induction on the step $i$.

\underline{Initialization ($i = 0$):}
Let $j$ be some fixed value in $\JSET(0)$. Since $\front_0$ is the connected component of $[U = \linfty]$ containing $\pinfty$, we have $\forall x \in \intbd \cavij{0}{j}, \linfty \not \in U(x)$ since by construction, we have:

$$\intbd \cavij{0}{j} \subseteq \cup_{j\in \JSET(0)} \intbd \cavij{0}{j} = \St{\front_0} \setminus \front_0.$$

Now, let us assume that we arbitrarily choose $x$ and $y$ in $\intbd \cavij{0}{j}$. Two cases are possible then: either $x$ and $y$ are adjacent, then $U(x) \subseteq U(y)$ or $U(y) \subseteq U(x)$, which means that they are of the same type relatively to $\linfty$ (both greater or both lower); or they are not adjacent, and we assume that $x$ and $y$ satisfy $U(x) \rhd \linfty $ and $ U(y) \lhd \linfty$. Thus, by Property~\ref{property.interiorboundary.connected}, there exists some path $\Pi$ of length $m$ in $\intbd  \cavij{0}{j}$ joining $x$ and $y$. Since $x = \Pi(0)$ and $y = \Pi(m)$ are assumed not to be of the same type, there exists $\ostar \in [0, m-1]$ satisfying that $\Pi(\ostar)$ and $\Pi(\ostar+1)$ are not of the same type, which is impossible as just seen before (either $U(\Pi(\ostar)) \subset U(\Pi(\ostar+1))$ or $U(\Pi(\ostar+1)) \subset U(\Pi(\ostar))$). Then, all the elements of $\intbd  \cavij{0}{j}$ are of the same type. Furthermore, since $\partial \cavij{0}{j} \subseteq P_0$, we have that $\forall x \in \partial \cavij{0}{j}, \lpar \in U(x)$.

\medskip

\underline{Heredity ($i \geq 1$):} we assume that at step $k \in [0,i-1]$, we have that for $j \in \JSET(k)$, for any $x \in \intbd \cavij{j}{k}$, $U(x) \rhd \levelCAV(\parentCAV(\cavij{j}{k}))$ (resp. for any $x \in \intbd \cavij{j}{k}$, $U(x) \lhd \levelCAV(\parentCAV(\cavij{j}{k}))$). Furthermore, for any $x \in \partial \cavij{j}{k}$, $U(x) \ni \levelCAV(\parentCAV(\cavij{j}{k}))$. We want to prove this property for $i$.

\medskip

Let $\cavij{i}{j}$ be a \PCAV at step $i > 0$, we can assume that it has a parent \PCAV $\PARENT$ with $\istar < i$ and $\jstar \in \JSET(\istar)$, where we have propagated the front using the level
$$\lpar = \levelCAV(\parentCAV(\cavij{j}{i}))$$
to cover:
$$\bigcup_{x \in \intbd \PARENT} \CC([U = \lpar] \cap,x) \cap \PARENT.$$
After this, we can deduce that the interior boundary of the \PCAV $\cavij{i}{j}$ satisfies:
$$\forall p \in \intbd \cavij{i}{j}, \ U(p) \neq \lpar,$$
that is, either $U(p)$ is greater than $\lpar$ or it is lower than $\lpar$. Two cases are then possible:

\begin{itemize}
    
    \item When $\cavij{i}{j}$ satisfies:
    $$ \Cl{\cavij{i}{j}} \subseteq \PARENT,$$
    then the (interior boundary of) $\cavij{i}{j}$ can be of positive or negative nature. As explained in the initialization, all the faces of the interior boundary are then of the same nature.
    
    \item When $\cavij{i}{j}$ satisfies:
    $$ \Cl{\cavij{i}{j}} \not \subseteq \PARENT,$$
    thus:
    $$\intbd \cavij{i}{j} \cap \intbd \PARENT \neq \emptyset,$$
    then we assume without constraint that $\PARENT$ is an upper \PCAV (the proof of the decreasing case is symmetrical). Thus two cases can occur:

    \begin{itemize}
        
        \item for the elements $p$ of $\intbd \cavij{i}{j} \cap \intbd \PARENT$, we did not have any propagation over $p$, which means that $U(p) \rhd \lpar$. 
        
        \item when $p$ belongs to $\intbd \cavij{i}{j} \setminus \intbd \PARENT$, propagation at the level $\lpar$ occurred until $p$ (excluded), which means that $U(p)$ is of type $\lhd \lpar$ or $\rhd \lpar$. Now, let $p^*$ be one element of $\intbd \cavij{i}{j} \cap \intbd \PARENT$, with $U(p^*) \rhd \lpar$. By connectivity of $\intbd \cavij{i}{j}$, we know that there exists a path joining $p$ and $p^*$, then by following the reasoning seen in the initialization, $U(p)$ is of type $\rhd \lpar$.

    \end{itemize}    
        
\end{itemize}

Concerning the proof that $\forall x \in \bd\cavij{i}{j}, \lpar \in U(x)$, either we are in the case:
$$ \Cl{\cavij{i}{j}} \subseteq \PARENT,$$
and then any $x$ of $\partial \cavij{i}{j}$ satisfies $U(x) \ni \lpar$, or we are in the case:
$$ \Cl{\cavij{i}{j}} \not \subseteq \PARENT,$$
and for any $h \in  \Cl{\cavij{i}{j}} \setminus \PARENT$, we look for the nearest parent $\PARENTDEUX$ of $\cavij{i}{j}$ which contains $h$, and then by Proposition~\ref{propo.polarity.cte} we obtain that $U(h) \supseteq [\levelCAV(\PARENTDEUX),\levelCAV(\cavij{i}{j}))]$ with:
$$\levelCAV(\PARENTDEUX) < \lpar = \levelCAV(\PARENT) < \levelCAV(\cavij{i}{j})),$$
and thus $\lpar \in U(h)$.}

\begin{Remark}
\label{remark.levels.cavities}
Let $\cav$ be some \PCAV of a \SBIPMAP $U$ not equal to $\CLOSEDDOM$. When this is an upper \PCAV:
\begin{align*}
\levelCAV(\cav) &= \sup\left\{\lambda \in \Reals \; ; \; \forall x \in \intbd(\cav), U(x) \rhd \lambda \right\}\\
&= \min\{ \lfloor U(x)\rfloor ; \forall x \in \intbd(\cav) \},
\end{align*}
and when this is a lower \PCAV:
\begin{align*}
\levelCAV(\cav) &= \inf\left\{\lambda \in \Reals \; ; \; \forall x \in \intbd(\cav), U(x) \lhd \lambda \right\}\\
&= \max\{ \lceil U(x)\rceil ; \forall x \in \intbd(\cav) \}.
\end{align*}
\end{Remark}

\begin{Property}
\label{P.is.connected.lbl}
Let $N$ be the (finite) number of iterations of the FPA. Then, for any value $i \in [0,N-1]$, the set $P_i$ is connected and $p_\infty \in P_i$.
\end{Property}

\IEEEproof{Let us proceed by induction on $i \geq 0$:
\begin{itemize}

\item \underline{Initialization:} $\PROPAG_1$ is connected by hypothesis, and $\pinfty$ belongs to $\PROPAG_1$.

\item \underline{Heredity:} Let us assume that $\PROPAG_i$ is connected and that $\pinfty$ belongs to $\PROPAG_i$ for some value $i \geq 1$. Then, $\PROPAG_{i + 1}$ is the union of $\PROPAG_i$, which is connected, and a set of connected components adjacent to $\PROPAG_i$. Thus, $\PROPAG_{i + 1}$ is connected and $\pinfty$ belongs to it.

\end{itemize}

By applying the induction, we obtain that the property is true for all the possible values of $i$ since $N$ is finite.}

\begin{Property}
The \PCAVS of a \SBIPMAP $U$ do not have any cavities.
\label{property.holes.have.no.hole}
\end{Property}

\IEEEproof{For sake of simplicity, let us denote by $\cavi{i}$ the set $\cup_{j \in \JSET(i)}\cavij{i}{j}$.

Let $\{H_l\}_{l\in [1,L]}$ the set of (non-empty) cavities of $\cavi{i}$. Now let us choose any fixed $k \in [1,L]$ and its corresponding $H_k$, cavity of some $\cavij{i}{j}$. By definition, $H_k \subset \sat(\cavi{i})$. Then, $\St{H_k} \subseteq \St{\sat(\cavi{i})} = \sat(\cavi{i})$ by Property~\ref{property.holes.are.open}. Then  $\St{H_k} \subseteq \sat(\cavi{i})$. The consequence is that: 
$$\St{H_k} \setminus \cup_{k \in [1,L]} H_k \subseteq \sat(\cavi{i}) \setminus \cup_{k \in [1,L]} H_k \subseteq \cavi{i}.$$

At the same time, $H_k \subseteq \CLOSEDDOM \setminus \cavi{i} \subseteq P_i$. Since $P_i$ is connected by Property~\ref{P.is.connected.lbl}, there exists some path $\Pi = (h,\dots,\pinfty)$ of length $m$ in $P_i$ joining $h \in H_k$ to $\pinfty \in P_i \setminus H_k$.

Now, let $\ostar$ be the value $\max_{o \in [1,m]} \{o \in [1,m] | \Pi(o) \in H_k\}$. Then, since $H_k$ is a closed set, we know that $\Pi(\ostar+1)$ belongs to $\St{H_k} \setminus H_k \subseteq \cavi{i} \subseteq \CLOSEDDOM \setminus P_i$, then $\Pi(\ostar+1)$ belongs (also) to $P_i$. We have a contradiction. The set $\cup_{l \in [1,L]}H_l$ is then empty.}

The following notation is depicted in Figure~\ref{fig:own_elements_cav}.

\begin{Notation}
Let $\cav$ be a \PCAV of $U$, then we define:
$$\properCAV{\cav} = \left(\bigcup_{x \in \intbd \cav} \CC([U = \levelCAV(\cav)],x)\right) \cap \cav.$$
\label{def.cav.star}
\end{Notation}

We recall that $\bigsqcup$ denotes the disjoint union operator. The following property can be observed in Figure~\ref{fig:own_elements_cav}.

\begin{Property}
\label{property.proper.elements.cavity}
Let $\cav$ be a \PCAV of $U$. We have the following equality:
$$\cav = \properCAV{\cav} \bigsqcup \left(\bigcup_{\cav' \in \childrenCAVITY(\cav)} \cav'\right),$$
or in other words:
$$\properCAV{\cav} = \cav \setminus \left(\bigcup_{\cav' \in \childrenCAVITY(\cav)} \cav'\right).$$
\end{Property}

\IEEEproof{Let $h$ be an element of some \PCAV $\cav$ of $U$. Two cases are possible. The first case corresponds to when $h$ belongs to some child $\cav'$ of $\cav$ and then any path starting from $h$ reaching $x \in \intbd(\cav) \cap [U = \levelCAV(\cav)]$ crosses the internal boundary of the direct child $\cav''$ of $\cav$ containing $\cav'$ on which $U$ is lower or greater than $\levelCAV(\cav)$ (see Proposition~\ref{propo.holes.have.uniform.contour.lbl}). The case corresponds to when there exists a path in $[U = \levelCAV(\cav)] \cap \cav$ joining $h$ and $x \in \intbd(\cav) \cap [U = \levelCAV(\cav)]$. Using Proposition~\ref{propo.cav}, we conclude the proof.}

\subsection{Mathematical properties of shapes}

\begin{Definition}
From now on, for any subset $E$ of $\CLOSEDDOM$, we will denote by $\Interior(E)$ the topological interior of $E$, that is, the greatest open set contained in $E$.
\end{Definition}

\subsubsection{Properties of shapes of a WC \SBIPMAP}

\begin{Proposition}[Shapes are open]
Let $\shape$ be some shape of $U$. Then, $\shape$ is an open set.
\label{shapes.are.open}
\end{Proposition}

\IEEEproof{An upper threshold set is the core~\cite{aubin2009set,najman2013discrete} image by the interval-valued map of an open interval $]a,+\infty[$, which is then open by upper semicontinuity of the span-based immersion $U$. Thus, its connected components are open. It follows that the saturations of these components, the shapes of $U$, are open by Proposition~\ref{saturation.openset.is.openset}. The same reasoning can be applied to lower shapes.}

\begin{Proposition}
Let $\CLOSEDDOM$ be a $n$-D hyper-rectangle in the $n$-D Khalimsky grid, and let $\pinfty$ be a $n$-face of $\CLOSEDDOM$ which belongs to $\partial \CLOSEDDOM$. When a digital set $\shape \subset \CLOSEDDOM$ is an open shape of a WC \SBIPMAP, then we have the following property:
$$\Interior(\sat(\partial \shape)) = \shape.$$
\label{proposition.shape.interior.sat.partial.shape}
\end{Proposition}

\IEEEproof{When we are in the case $\shape = \CLOSEDDOM$, we obtain easily:
$$\sat(\shape) = \CLOSEDDOM \setminus \connectedcomponent(\CLOSEDDOM \setminus \partial \CLOSEDDOM,\pinfty) = \CLOSEDDOM.$$
When $\shape \subsetneq \CLOSEDDOM$, we can use Proposition~\ref{propo.sat.inclusion} and we obtain:
$$\sat(\Cl{\shape}) = \sat(\partial \Cl{\shape}) = \sat(\Cl{\Cl{\shape}} \setminus \Interior(\Cl{\shape})).$$
Thanks to the idempotency of the closure operator and to Proposition~\ref{propo.plain.maps.regular} (regularity of the open shape $\shape$):
$$\sat(\Cl{\shape}) = \sat(\Cl{\shape} \setminus \shape) = \sat(\partial \shape).$$
Thanks to Remark~\ref{remark.shapes.WC}, $\shape$ is WC, and then by Proposition~\ref{propo.satCC=CCsat}:
$$\sat(\partial \shape) = \sat(\Cl{\shape}) = \Cl{\sat(\shape)} = \Cl{\shape}.$$
This way, $$\Interior(\sat(\partial \shape)) = \Interior(\Cl{\shape}) = \shape$$ by regularity of $\shape$.}

\medskip

Let us now assume that the studied function $U$ is WC.

\begin{Notation}
Let $\shape$ be a shape of a WC \SBIPMAP $U$. We define the following set:
$$\nextSHAPE(\shape) = \left\{\shape' \in \ToS(U) \; ; \; \shape = \parentSHAPE(\shape')\right\}.$$
\end{Notation}

\begin{Notation}
Let $\shape$ be a shape of a WC \SBIPMAP $U$. We denote by $\childrenSHAPE(\shape)$ the set of children of $\shape$ in $\ToS(U)$.
\end{Notation}

We obtain then easily the following proposition.

\begin{Proposition}
Let $U$ be a WC \SBIPMAP on $\CLOSEDDOM$. For any shape $\shape$ of $U$, we have the following equality:
$$\bigcup_{\shape' \in \nextSHAPE(\shape)} \shape'  = \bigcup_{\shape' \in \childrenSHAPE(\shape)} \shape'.$$
\label{propo.tos}
\end{Proposition}

\begin{figure}[htb!]
\begin{center}
\includegraphics[width=0.88\linewidth]{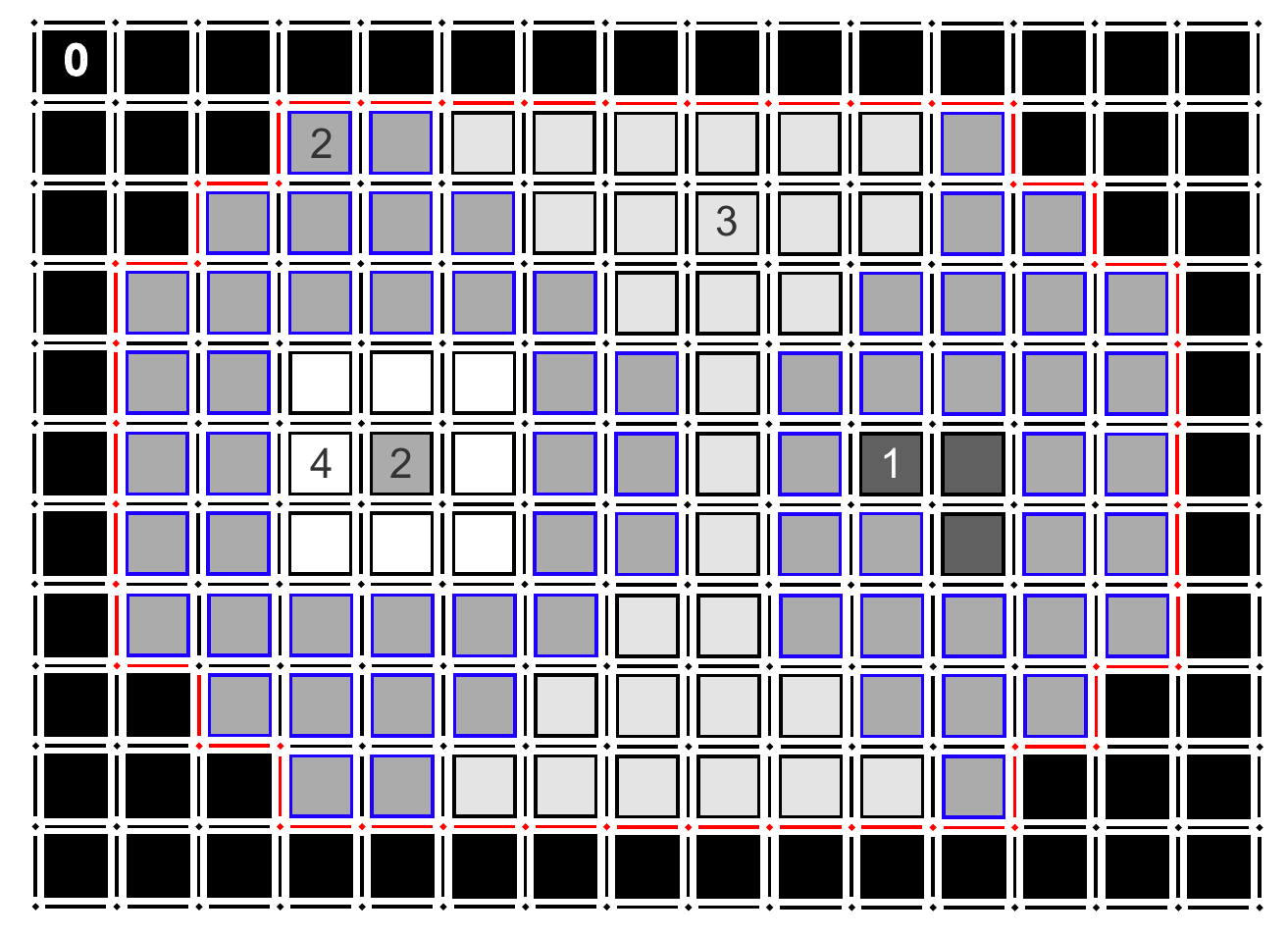}
\end{center}
\caption{Starting from the upper shape $\shape = \sat([U \rhd 0.5])$ whose contour is colored in red, we compute $\properSHAPE{\shape}$ depicted by squares with blue contours; remark that $\levelSHAPE(\shape)$ is equal to $2$ since it is the minimal value that can be found in the internal boundary.}
\label{fig:own_elements_shape}
\end{figure}

\begin{Definition}[Levels of shapes]
We define:
$$\levelSHAPE(\CLOSEDDOM) = \linfty.$$
For any shape $\shape$ of $U$ different from $\CLOSEDDOM$, we define when this shape is an upper shape:
\begin{align*}
\levelSHAPE(\shape) &= \sup\left\{\lambda \in \Reals \; ; \; \forall x \in \intbd(\shape), U(x) \rhd \lambda \right\}\\
&= \min\{\lfloor U(x) \rfloor ;x \in \intbd(\shape)\},
\end{align*}
and when this shape is a lower shape, then:
\begin{align*}
\levelSHAPE(\shape) &= \inf\left\{\lambda \in \Reals \; ; \; \forall x \in \intbd(\shape), U(x) \lhd \lambda \right\}\\
&= \max\{\lceil U(x) \rceil ;x \in \intbd(\shape)\}.
\end{align*}
\end{Definition}

This definition is depicted in Figure~\ref{fig:own_elements_shape}.

\begin{Notation}
Let $U$ be some WC \SBIPMAP and $\ToS$ its tree of shapes. Then for any shape $\shape$ of $U$ different from the root $\CLOSEDDOM$ of $\ToS$, we denote by $\parentSHAPE(\shape)$ the direct parent of $\shape$ in $\ToS$.
\end{Notation}

\begin{Proposition}
\label{propo.upper.increasing}
Let $\shape$ be some shape of some WC \SBIPMAP $U$ with $\shape \neq \CLOSEDDOM$, then when $\shape$ is an upper shape, it satisfies that $\levelSHAPE(\shape) > \levelSHAPE(\parentSHAPE(\shape))$. Conversely, when $\shape$ is a lower shape, it satisfies that $\levelSHAPE(\shape) < \levelSHAPE(\parentSHAPE(\shape))$.
\end{Proposition}

\IEEEproof{Let us treat the upper case only, the other case being symmetrical. Let us assume that $\shape$ is an upper shape different from $\CLOSEDDOM$, then there exists some connected component $\Gamma \in \CC([U \rhd \levelSHAPE(\shape) - \varepsilon])$ with $\varepsilon \rightarrow 0^+$ such that $\sat(\Gamma) = \shape$. However, if we assume now that $\levelSHAPE(\parentSHAPE(\shape)) > \levelSHAPE(\shape)$, then there exists some face $x \in \parentSHAPE(\shape)$ which is neighbor of $\Gamma$ and which satisfies $U(x) \rhd \levelSHAPE(\shape)$. By construction of $\Gamma$, it contains $x$, and then we obtain that $\sat(\Gamma) \supsetneq \shape$, which is a contradiction. However, the case $\levelSHAPE(\parentSHAPE(\shape)) = \levelSHAPE(\shape)$ is impossible, then $\levelSHAPE(\parentSHAPE(\shape)) < \levelSHAPE(\shape)$.}

\begin{Proposition}
\label{propo.sharing.implies.same.type}
Let $\shape,\shape'$ be two shapes different from $\CLOSEDDOM$ of a WC \SBIPMAP $U$ with $\shape' = \parentSHAPE(\shape)$ and with $\cl(\shape) \not \subseteq \shape'$. Then, $\shape$ and $\shape'$ are both upper shapes or both lower shapes.
\end{Proposition}

\IEEEproof{Let us treat the case where $\shape$ is an upper shape (he lower case follows a same reasoning). We assume that $\shape'$ is a lower shape, then for any $x \in \intbd \shape'$, we have $U(x) \lhd \levelSHAPE(\shape') + \varepsilon$ with $\varepsilon \rightarrow 0^-$. Since $\cl(\shape) \not \subseteq \shape'$, then there exists a face $h$ in $\cl(\shape)$ which does not belong to $\shape'$, then there exists $x$ in the closure of $h$ which belongs to:
$$\intbd \shape \cap \intbd \shape',$$
which leads at the same time to $U(x) \rhd \levelSHAPE(\shape) - \varepsilon$ with $\varepsilon \rightarrow 0^+$ and to $U(x) \lhd \levelSHAPE(\shape') + \varepsilon'$ with $\varepsilon' \rightarrow 0^-$. Then, we obtain:

$$\levelSHAPE(\shape) - \varepsilon \lhd U(x) \lhd \levelSHAPE(\shape') + \varepsilon',$$
which means that $\levelSHAPE(\shape') > \levelSHAPE(\shape)$, then by Proposition~\ref{propo.upper.increasing}, $\shape$ is a lower shape. We obtain a contradiction. This concludes the proof.}

\begin{Proposition}
\label{propos.upper.is.rhd}
Let $\shape$ be a shape of a WC \SBIPMAP $U$ different from $\CLOSEDDOM$. When $\shape$ is an upper shape:
$$\forall \lambda \in [\levelSHAPE(\parentSHAPE(\shape)),\levelSHAPE(\shape)[, \forall x \in \intbd \shape, \ U(x) \rhd \lambda,$$
and
$$\forall \lambda \in [\levelSHAPE(\parentSHAPE(\shape)),\levelSHAPE(\shape)], \forall x \in \bd \shape, \ U(x) \ni \lambda,$$
and $\shape$ is a lower shape:
$$\forall \lambda \in ]\levelSHAPE(\shape),\levelSHAPE(\parentSHAPE(\shape))], \forall x \in \intbd \shape, \ U(x) \lhd \lambda,$$
and
$$\forall \lambda \in [\levelSHAPE(\shape),\levelSHAPE(\parentSHAPE(\shape))], \forall x \in \intbd \shape, \ U(x) \ni \lambda.$$
\end{Proposition}

\IEEEproof{Let us treat the upper case only (the other case can be deduced by symmetry). By Proposition~\ref{propo.upper.increasing}, we know that:
$$\levelSHAPE(\parentSHAPE(\shape)) < \levelSHAPE(\shape).$$
Now, let us prove that for any $x \in \intbd \shape$, we have that $U(x) \rhd \levelSHAPE(\shape) - \varepsilon$ with $\varepsilon \rightarrow 0^+$. For this aim, we assume that there exists at least one face $h \in \intbd(\shape)$ such that $U(h) \not \rhd \levelSHAPE(\shape) - \varepsilon$. Since $\shape$ is an upper shape, there exists a connected component $\Gamma$ of $[U \rhd \levelSHAPE(\shape) - \varepsilon]$ such that $\sat(\Gamma) = \shape$. However, $h$ does not belong to $\Gamma$. Then, $h$ belongs to $\CLOSEDDOM \setminus \Gamma$. Also, $h \in \intbd(\shape)$, then there exists some path $\Pi$ joining $\pinfty$ to $h$ in $\CLOSEDDOM \setminus \shape$ (since $\CLOSEDDOM \setminus \shape$ is connected). Then, $h$ belongs to $\CC(\CLOSEDDOM \setminus \Gamma,\pinfty)$, and then $h \not \in \sat(\Gamma)$, that is, $\intbd(\shape) \not \subseteq \shape$, which is a contradiction. Then all the faces $x$ of $\intbd$ satisfy the property that $U(x)$ is greater than $\levelSHAPE(\shape)$. Concerning the property that for any $x \in \partial \shape$, $U(x)$ contains $\levelSHAPE(\shape)$, it is simply due to the properties of the span-based immersion:
\begin{itemize}

    \item when $\cl(\shape) \subseteq \parentSHAPE(\shape)$, each face $x$ of $\partial \shape$ is neighbor of a $n$-face $h_{in}$ of $\intbd \shape$ whose value is $U(h_{in}) = {v}$ with $v \geq \levelSHAPE(\shape)$ and of an $n$-face $h_{out}$ in $\properSHAPE{\parentSHAPE(\shape)}$ with $U(h_{out}) = \levelSHAPE(\parentSHAPE(\shape))$, then it contains $$[\levelSHAPE(\parentSHAPE(\shape)),v],$$ and then $\levelSHAPE(\shape)$,

    \item when $\cl(\shape) \not \subseteq \parentSHAPE(\shape)$, we have two possible cases. Either we have $x \in \partial \shape$ which belong to $\parentSHAPE(\shape)$ and we can reason as in the previous case, or $x \in \partial \shape$ does not belong to the (direct) parent $\parentSHAPE(\shape)$. In this last case, we look for the nearest parent $\shape'$ in $\ToS(U)$ which contains $x$. Since it will be an upper shape by Proposition~\ref{propo.sharing.implies.same.type}, we will have that $\levelSHAPE(\shape')$ is lower than $\levelSHAPE(\shape)$. Now, let us define $v$ as the supremum of (the lower bounds of) $U$ on the $n$-faces of $\cl(x)$; we will obtain $v \geq \levelSHAPE(\shape)$. We will obtain that $U(x)$ contains:
    $$[\levelSHAPE(\shape'),v]$$ and then $\levelSHAPE(\shape)$.

\end{itemize}

This concludes the proof.}

\begin{Notation}
\label{notation.proper.shape}
Let $\shape$ be a shape of $U$. We will define:
$$\properSHAPE{\shape} = \shape \setminus \left(\bigcup_{\shape' \in \childrenSHAPE(\shape)} \shape'\right).$$
\end{Notation}

This notation is depicted in Figure~\ref{fig:own_elements_shape}.

\begin{Remark}
Let $\shape$ be a shape of $U$. Then, any $n$-face $h$ of $\properSHAPE{\shape}$ satisfies:
$$U(h) = \left\{\levelSHAPE(\shape)\right\}.$$
\end{Remark}

\begin{Proposition}[Separation property of shape boundaries]
Let $\shape$ be an (open) shape of a \SBIPMAP $U$ defined on $\CLOSEDDOM$. Let choose an operator in $\{\intbd,\partial\}$. We say that $\BOUNDARY (\shape)$ separates $\CLOSEDDOM$ in the sense that when we call \emph{exterior component} the set:
$$\EspaceExterieur := \CC(\CLOSEDDOM \setminus \BOUNDARY (\shape),\pinfty)$$
and \emph{interior component} the set:
$$\EspaceInterieur = \Interior(\sat(\BOUNDARY(\shape))),$$
then we can assert that any path in $\CLOSEDDOM$ going from the interior component to the exterior component (or the converse) will intersect $\BOUNDARY (\shape)$. Furthermore, we obtain the following partition of $\CLOSEDDOM$: 
$$\CLOSEDDOM = \EspaceExterieur \sqcup \BOUNDARY(\shape) \sqcup \EspaceInterieur.$$
\label{proposition.interior.boundary.sperates}
\end{Proposition}

\IEEEproof{Let us begin with the case $\BOUNDARY = \partial$. Since $\shape$ is open, we know by Proposition~\ref{proposition.shape.interior.sat.partial.shape} that:
$$\EspaceInterieur = \Interior(\sat(\partial \shape))) = \shape,$$
and at the same time we obtain that the exterior component is equal to:
$$\EspaceExterieur = \CLOSEDDOM \setminus \cl(\shape),$$
which is due to the fact that $\sat(\partial \shape) = \cl(\shape)$ (see the proof of Proposition~\ref{proposition.shape.interior.sat.partial.shape}). Then, every path $\Pi$ going from $\EspaceExterieur$ to $\shape$ satisfies that it starts from $a \in \EspaceExterieur$ to $b \in \shape$ in $\CLOSEDDOM$. Let $\Pi(\istar)$ be the last element of $\Pi$ which is in $\EspaceExterieur$, then $\Pi(\istar + 1)$ belongs to $\shape$ or to $\partial \shape$. Since $\EspaceExterieur$ is open, $\Pi(\istar+1)$ does not belong to $\st(\Pi(\istar)$, otherwise it belongs to $\EspaceExterieur$. Then, $\Pi(\istar+1)$ belongs to $\cl(\Pi(\istar)) \setminus \EspaceExterieur$, and then to $$\cl(\EspaceExterieur) \setminus \EspaceExterieur = \partial \EspaceExterieur = \partial \shape.$$ The same reasoning applies when we start from $b$ in $\shape$ and then every path $\Pi$ from $\EspaceExterieur$ to $\EspaceInterieur$ intersects $\partial \shape$.

\medskip

Let us treat now the case $\BOUNDARY = \intbd$. We recall by Proposition~\ref{proposition.reformulation.interior.boundary} that:
$$\intbd \shape = \st(\KHAL{n} \setminus \shape) \cap \st(\shape)$$ or after simplification:
$$\intbd \shape = \st(\CLOSEDDOM \setminus \shape) \cap \shape.$$
Now, let $\Pi$ be a path in $\CLOSEDDOM$ starting from a belonging to:
$$\EspaceExterieur = \CLOSEDDOM \setminus \shape,$$
and going to $b$ belonging to:
$$\EspaceInterieur = \shape \setminus \intbd \shape.$$
Then, there exists $\istar$ such that $\Pi(\istar)$ is the last element of $\Pi$ which does not belong to $\shape$. It satisfies $\Pi(\istar + 1) \in \st(\Pi(\istar)) \setminus \{\Pi(\istar)\}$, then $\Pi(\istar + 1)$ belongs to $\shape$ and to $\st(\CLOSEDDOM \setminus \shape)$, then to $\intbd \shape$. Conversely, if we start from $b$, we denote by $\istar$ the index of $\Pi$ such that $\Pi(\istar)$ is the last element in $\shape$. Then, $\Pi(\istar+1)$ belongs to $\EspaceExterieur$, then $\Pi(\istar)$ belongs to $\st(\CLOSEDDOM \setminus \shape) \cap \shape = \intbd \shape$ and $\Pi$ intersects $\intbd \shape$.}

\medskip

\begin{figure}[htb!]
\begin{center}
\includegraphics[width=0.88\linewidth]{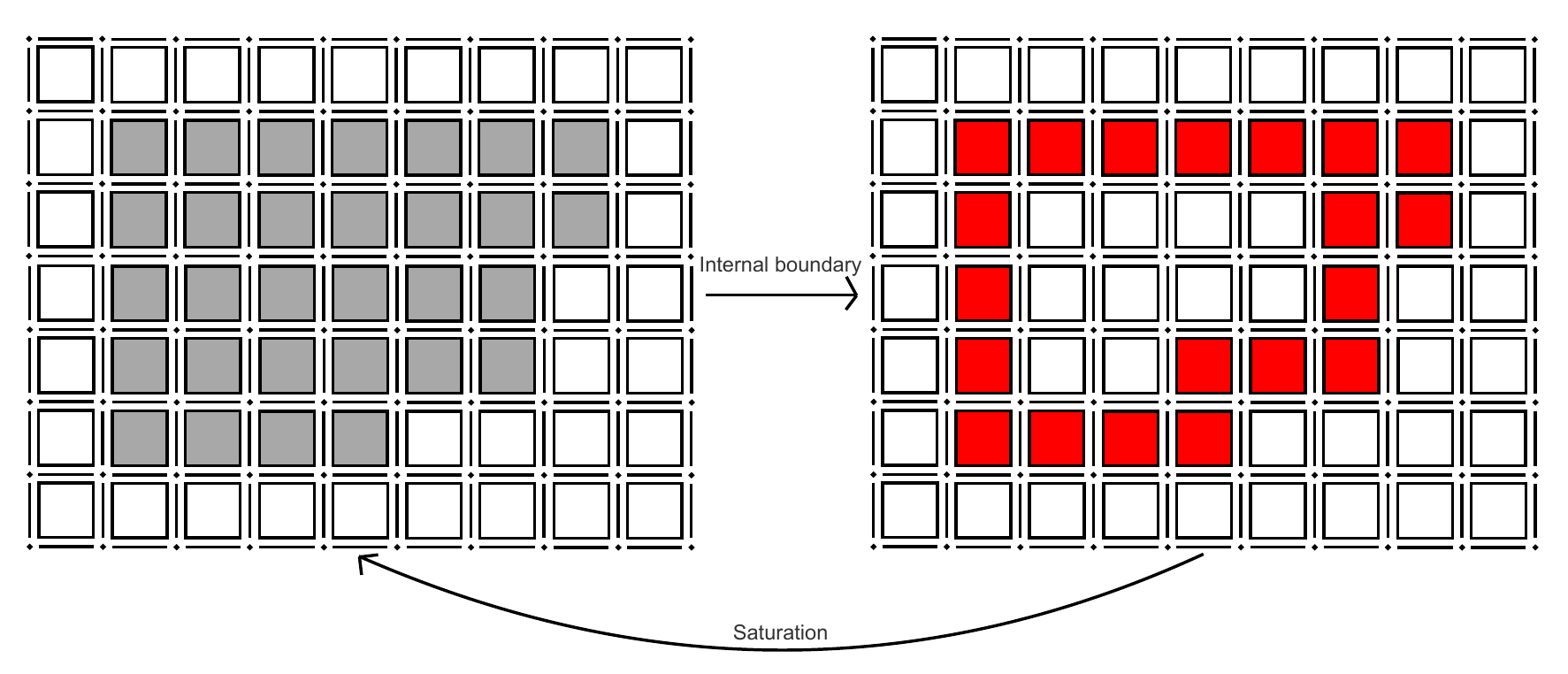}
\end{center}
\caption{The saturation of the internal boundary of a shape of a WC \SBIPMAP is equal to this same shape}
\label{fig:prop_internal}
\end{figure}

\begin{Proposition}
Let $U$ be some WC \SBIPMAP and $\shape$ one of its shapes. We have the following property:
$$\sat(\intbd(\shape)) = \shape.$$
\label{propo.sat.bdint.shape.equals.shape}
\end{Proposition}

\IEEEproof{This proof is depicted in Figure~\ref{fig:prop_internal}. Using the notations of the proof of Proposition~\ref{proposition.interior.boundary.sperates}, we obtain:
$$\sat(\intbd(\shape)) = \CLOSEDDOM \setminus \CC(\CLOSEDDOM \setminus \intbd(\shape),\pinfty),$$
and the two (separated) connected components of $\CLOSEDDOM \setminus \intbd(\shape)$ are $\EspaceExterieur$ and $\EspaceInterieur$. However, only $\EspaceExterieur$ contains $\pinfty$. Thus, 
$$\CC(\CLOSEDDOM \setminus \intbd(\shape),\pinfty) = \EspaceExterieur,$$
and $\sat(\intbd(\shape)) = \CLOSEDDOM \setminus \EspaceExterieur = \EspaceInterieur \sqcup \intbd(\shape) = \shape$.
This concludes the proof.}
\medskip

\begin{Proposition}
Let $\shape$ be some shape of $U$, then we have the remarkable properties (as depicted in Figure~\ref{fig:own_elements_shape}):
$$\properSHAPE{\CLOSEDDOM} = \bigcup_{x \in \border \CLOSEDDOM} \CC\left([U = \linfty],x\right).$$
and when $\shape \neq \CLOSEDDOM$:
$$\properSHAPE{\shape} = \left(\bigcup_{x \in \intbd(\shape)} \CC\left([U = \levelSHAPE(\shape)],x\right)\right) \cap \shape.$$
\label{proposition.proper.shape}
\end{Proposition}

\IEEEproof{The case of $\CLOSEDDOM$ is obvious. We can observe that this proposition results from the fact that when we have some face $h$ in some upper shape $\shape$ of $U$ (then by Proposition~\ref{propos.upper.is.rhd} it is a shape of type $\rhd$), four cases are possible:

\begin{itemize}
    
    \item[(1)] $\levelSHAPE(\shape) \in U(h)$ and there exists some path joining $h$ to some $x \in \intbd \shape$ in $[U = \levelSHAPE(\shape)]$. In this case, $h$ belongs to $\properSHAPE{\shape}$.
    
    \item[(2)] $\levelSHAPE(\shape) \in U(h)$ but there does not exist any path joining $h$ to some $x \in \intbd \shape$ in $[U = \levelSHAPE(\shape)]$, then there exists two shapes $\shape',\shape'' \in \childrenSHAPE(\shape)$ of $U$ such that $\levelSHAPE(\shape') \neq \levelSHAPE(\shape)$, $\levelSHAPE(\shape'') = \levelSHAPE(\shape)$, and $\shape'' \subset \shape' \subset \shape$. In this case, $h$ belongs to $\shape''$ which belongs to the children of $\shape$, and then $h \not \in \properSHAPE{\shape}$.
    
    \item[(3)] $U(h) \lhd \levelSHAPE(\shape)$, then $\intbd \shape$ (where $U$ is $\rhd$ $\levelSHAPE(\shape) - \varepsilon$ with $\varepsilon \rightarrow 0^+$) separates $\CLOSEDDOM$ by Proposition~\ref{proposition.interior.boundary.sperates} so that we obtain:
    $$\CC\left([U \lhd \levelSHAPE(\shape)],h\right) \subseteq \sat(\intbd \shape),$$ then by Proposition~\ref{propo.sat.increasing} and Proposition~\ref{propo.sat.bdint.shape.equals.shape}:
    $$\shape' = \sat(\CC\left([U \lhd \levelSHAPE(\shape)],h\right)) \subseteq \shape.$$
    Since $\shape'$ is a lower shape different from $\CLOSEDDOM$, $\shape' \neq \shape$, and then $\shape' \subsetneq \shape$. Thus, $h$ belongs to $\shape'$ which belongs to the children of $\shape$, and then $h \not \in \properSHAPE{\shape}$.
    
    \item[(4)] $U(h) \rhd \levelSHAPE(\shape)$, then $\bd \shape$ (where $U$ contains $\levelSHAPE(\shape)$) separates $\CLOSEDDOM$ by Proposition~\ref{proposition.interior.boundary.sperates} so that we obtain:
    $$\CC\left([U \rhd \levelSHAPE(\shape)],h\right) \subseteq \Interior(\sat(\bd \shape)),$$
    which means that $\shape' = \sat(\CC\left([U \rhd \levelSHAPE(\shape)],h\right)) \subseteq \shape$, and then there exists a child $\shape'$ of $\shape$ satisfying $h \in \shape'$, and then $h \not \in \properSHAPE{\shape}$.
    
\end{itemize}
Finally, $h$ belongs to $\properSHAPE{\shape}$ iff it belongs to $$\left(\bigcup_{x \in \intbd(\shape)} \CC\left([U = \levelSHAPE(\shape)],x\right)\right) \cap \shape,$$ which proves the assertion in the upper case. The same reasoning applies in the lower case. This concludes the proof. }

\subsection{Properties of cavities that are shapes}

\begin{Proposition}
\label{proposition.levels}
Let $U$ be a \SBIPMAP, and let $\BOTH$ be a subset of $\CLOSEDDOM$ which is at the same time a shape of $U$ and a \PCAV of $U$. Then $\levelCAV(\BOTH) = \levelSHAPE(\BOTH)$.
\end{Proposition}

\IEEEproof{It is a direct consequence of Remark~\ref{remark.levels.cavities}.}

\begin{Proposition}
\label{proposition.propersets}
Let $U$ be a \SBIPMAP, and let $\BOTH$ be a subset of $\CLOSEDDOM$ which is at the same time a shape of $U$ and a \PCAV of $U$. Then $\properCAV{\BOTH} = \properSHAPE{\BOTH}$.
\end{Proposition}

\IEEEproof{Let $\BOTH$ be a \PCAV and a shape of $U$. Then, thanks to Proposition~\ref{proposition.levels} and to Proposition~\ref{proposition.proper.shape}:

$$
\begin{array}{rcl}
\properCAV{\BOTH} & = & \left(\bigcup_{x \in \intbd(\BOTH)} \CC\left([U = \levelCAV(\BOTH)],x\right)\right) \cap \BOTH,\\
& = &\left(\bigcup_{x \in \intbd(\BOTH)} \CC\left([U = \levelSHAPE(\BOTH)],x\right)\right) \cap \BOTH,\\
& = & \properSHAPE{\BOTH}
\end{array}
$$

The proof is done.}

\subsection{Our main theorem}

\begin{Theorem}
Let $U$ be a well-composed \SBIPMAP defined on a unicoherent domain $\CLOSEDDOM$. The \PCAV hierarchy computed by the FPA is equal to the tree of shapes of $U$. In other words, our FPA algorithm computes the tree of shapes of $U$.
\label{th.tos=cavities}
\end{Theorem}

\IEEEproof{Let $\BOTH$ be a subset of $\CLOSEDDOM$ which is at the same time a \PCAV and a shape of $U$. Thanks to Proposition~\ref{proposition.propersets}, we obtain that :
$$\properCAV{\BOTH} = \properSHAPE{\BOTH},$$
which leads by Property~\ref{property.proper.elements.cavity} to:
$$\BOTH \setminus \left(\bigcup_{\cav \in \childrenCAVITY(\BOTH)} \cav\right) = \properSHAPE{\BOTH}.$$
Using Notation~\ref{notation.proper.shape}, we obtain:
$$\bigcup_{\cav \in \childrenCAVITY(\BOTH)} \cav = \bigcup_{\shape \in \childrenSHAPE(\BOTH)} \shape,$$
which can be reformulated by Proposition~\ref{propo.tos} and Proposition~\ref{propo.cav}:
$$\bigcup_{\cav \in \nextCAV(\BOTH)} \cav = \bigcup_{\shape \in \nextSHAPE(\BOTH)} \shape.$$

\medskip

Furthermore, the elements of $\nextCAV(\BOTH)$ are \PCAVS, then open by Property~\ref{property.holes.are.open}, and connected by definition. The elements of $\nextSHAPE(\BOTH)$ are shapes, then open by Proposition~\ref{shapes.are.open}, and connected by Proposition~\ref{propo.sat.connected.is.connected}.

\medskip

The consequence is that:
\begin{align}
\nextCAV(\BOTH) & = \CC\left( \bigcup_{\cav \in \nextCAV(\BOTH)} \cav \right)\\
&= \CC\left( \bigcup_{\shape \in \nextSHAPE(\BOTH)} \shape \right)\\
&= \nextSHAPE(\BOTH).
\end{align}

\medskip

We obtain thus directly that:
$$\HierarchyCavities = \ToS$$
since their root is the same set $\CLOSEDDOM$.}

\begin{Coro}
By following the procedure described in Figure~\ref{fig:summarycomputation}, the tree of \PCAVS computed on $u$ is the tree of shapes of $u$.
\label{coro.tos}
\end{Coro}

\section{Discussion about the tree of shapes}

In Corollary~\ref{coro.tos}, we assert that the tree of cavities computed by our algorithm is the tree of shapes of $u$. In fact, when the input image $u$ is already \WC, the immersion $U$ is \WC, and thus no topological reparation of $U$ is needed to obtain the property that the output of our algorithm (before the emersion) will be the tree of shapes of $U$. After the emersion, since it is used to remove the additional faces (added during the immersion step), we understand easily that the tree of cavities of $U$ becomes \textquote{the} tree of shapes of $u$.

\medskip

However, when the initial image $u$ is not \WC, we have to make it \WC using some topological reparation algorithm, based for example on a min-interpolation (simulating the $2n$-connectivity for the ones), or a max-interpolation (simulating the $3^n-1$-connectivity for the ones), or even our $n$-D self-dual topological reparation (considering different connectivities depending on the context~\cite{boutry2019make}). Due to this choice that has to be made, there exist several tree of shapes, depending on the choice that has been made about the \WC interpolation we have chosen before applying our algorithm.

\medskip

Another important remark is that the algorithm we present here is thus deterministic: assuming we have fixed the \WC interpolation that we used at the beginning of the algorithm, whatever the random propagation during the sort algorithm is (several sequences of levels are possible during the propagation depending on if we increase or decrease the level), the final output is always the tree of shapes. So the algorithm we present here is deterministic. Intuitively, this determinism is justified by the fact that depending on when we increase or decrease the current level, we propagate in different areas of the domain of the image, thus increasing or decreasing has no influence on the final computation.

\section{Complexity Analysis}
\label{ssec:complexity.analysis}

Complexity analysis of the algorithm presented here is trivial.  The immersion, canonicalization, and emersion are linear.  The modified union-find (once augmented with tree balancing, i.e., union-by-rank) is quasi-linear when values of the input image $u$ have a low quantization (typically 12~bits or less). Last, the time complexity of the sorting step is governed by the use hierarchical queue: it is linear with low quantized data\footnote{Formally the sorting step has the pseudo-polynomial O($k \, n$) complexity, $k$
  being the number of different gray values.  Though, since we
  consider low bit-depths data, $k$ shall only be considered as a
  complexity multiplicative factor.}.  Eventually, we obtain a quasi-linear algorithm.

\section{Conclusion}
\label{sec:conclusion}

We have proven in this paper that the tree computed in the algorithm provided in~\cite{geraud.2013.ismm} is indeed the expected tree of shapes. We also know that the tree of shapes is a purely self-dual representation of a cubical image when we work with self-dual well-composed interpolations. For this reason, we propose to focus in the future on the proof that the interpolation provided in~\cite{boutry.15.ismm} is self-dual.
\section*{Acknowledgments}

The authors would like to thank Michel Couprie, Pascal Monasse, and Yongchao Xu for fruitful discussions about this topic.

\bibliographystyle{IEEEtran}
\bibliography{article} %

\end{document}